\documentclass[a4paper,11pt]{article}
\usepackage{pos}

\newcommand{\nn}{\nonumber}
\newcommand{\mrm}[1]{\mbox{$\mathrm{#1}$}}

\title{\boldmath ($0,6$) AdS$_3$/CFT$_2$ and surface defects}

\author*[a,b]{Yolanda Lozano}
\author[c]{Anayeli Ram\'irez}

\affiliation[a]{Department of Physics, University of Oviedo\\ Avda. Calvo Sotelo s/n, 33007 Oviedo, Spain}

\affiliation[b]{Instituto Universitario de Ciencias y Tecnolog\'ias Espaciales de Asturias (ICTEA),\\
	Calle de la Independencia 13, 33004 Oviedo, Spain}

\affiliation[b]{Instituut voor Theoretische Fysica, KU Leuven,
	\\
	Celestijnenlaan 200D, 3000 Leuven, Belgium}

\emailAdd{ylozano@uniovi.es}
\emailAdd{mariaanayeli.ramirezortiz@kuleuven.be}

\abstract{We explore a new class of AdS$_3$ solutions in massive type IIA supergravity preserving $\mathcal{N} = (0,6)$ supersymmetry and realising an $\mathfrak{osp}(6|2)$ superconformal algebra. These solutions exhibit an SO(6)-symmetric internal space constructed from a $\mathbb{CP}^3$, and are fully specified by a single cubic function controlling the fluxes and warping.   
We propose a brane box configuration underlying the solutions from which we construct a two-dimensional quiver gauge theory whose anomaly structure and central charge we analyse, and from which we can realise Seiberg-like dualities as large gauge transformations. The brane box configuration suggests an interpretation of the solutions as dual to surface defects within the ABJ(M) theory. Our findings provide a concrete setting for exploring holography beyond the ABJM vacuum. Remarkably, no explicit field theories are currently known to realise $\mathcal{N} = (0,6)$ supersymmetry in two dimensions, making our setup a promising and largely unexplored direction for field-theoretic investigations. }

\FullConference{
}

\begin{document}
\maketitle
\section{Introduction}
\label{sec:intro}

The emergence of $\mathrm{AdS}_3$ geometries in the near-horizon limit of extremal black strings and black holes has long been recognised as a powerful window into the microscopic structure of quantum gravity. Unlike their higher-dimensional counterparts, $\mathrm{AdS}_3$ spaces uniquely support an infinite-dimensional conformal symmetry, allowing for unparalleled control over the structure of the dual two-dimensional CFT. This renders $\mathrm{AdS}_3$/CFT$_2$ the most tractable instance of holography, and a prime setting for precision tests of the holographic principle in strongly coupled regimes where stringy and gravitational degrees of freedom are deeply intertwined. Among its many successes, the ability of this correspondence to reproduce the Bekenstein--Hawking entropy of supersymmetric black holes \cite{Strominger:1996sh, Kraus:2005vz} stands out as one of the most compelling validations of the microscopic foundations of holography \cite{Brown:1986nw, Witten:2007kt, Guica:2008mu}.

Beyond its foundational role in early holographic developments, $\mathrm{AdS}_3$ continues to play a central part in modern string theory. It arises naturally as the near-horizon geometry of a broad class of intersecting brane configurations and flux compactifications \cite{Youm:1999zs, Bachas:2000fr, Gauntlett:2006af, Lozano:2019zvg, Couzens:2017nnr, Faedo:2020nol, Lozano:2019emq, Dibitetto:2018ftj, Lozano:2016wrs, Macpherson:2022sbs, Couzens:2021tnv, Passias:2020ubv, Couzens:2019mkh, Couzens:2019iog, Eberhardt:2017uup, Datta:2017ert, Lozano:2020txg, Eberhardt:2018ouy, Couzens:2022agr, Dibitetto:2018gtk}
, including those with reduced supersymmetry and intricate global symmetry structures. In particular, $\mathrm{AdS}_3$ backgrounds often encode the low-energy dynamics of defects, interfaces, or wrapped branes in higher-dimensional field theories.

This connection has led to the study of defect conformal field theories (dCFTs), in which a lower-dimensional conformal theory arises on the worldvolume of a defect embedded in a higher-dimensional parent theory \cite{Karch:2000gx, DeWolfe:2001pq}. Depending on whether conformal symmetry is partially preserved or completely broken, one obtains either a defect CFT or a defect-induced RG flow. Holographically, such systems are typically realised either via probe branes wrapping $\mathrm{AdS}$ submanifolds or through fully backreacted solutions when the defect density becomes large --signalling the emergence of a new, lower-dimensional holographic dual \cite{Aharony:2011yc, D'Hoker:2006uv, Chiodaroli:2011nr, DHoker:2007hhe, Faedo:2020nol}. An alternative viewpoint interprets these solutions as encoding the dynamics localised on the defect itself. In this picture, position-dependent couplings break part of the ambient conformal symmetry, leading to distinctive features such as non-vanishing one-point functions and the appearance of displacement operators \cite{Billo:2016cpy, Kobayashi:2018ezm}. In string theory, these effects naturally arise in intersecting brane setups, where the defect emerges at the interface between higher- and lower-dimensional branes. The resulting supergravity backgrounds exhibit a warped structure with an $\mathrm{AdS}$ slicing, effectively encoding the defect in the internal space geometry \cite{DHoker:2007zhm, Estes:2012nx, Lozano:2016wrs, Couzens:2022agr}. When the internal manifold is non-compact, this correspondence becomes particularly transparent: the non-compact directions reflect the UV origin of the defect theory, while the $\mathrm{AdS}$ factor captures its IR behaviour. These backgrounds often preserve remnants of higher-dimensional symmetries, offering a powerful framework to study the interplay between defect dynamics and symmetry breaking.

Motivated by the sustained progress in understanding the supersymmetry constraints of $\mathrm{AdS}_3$ backgrounds~\cite{Dibitetto:2018ftj, Passias:2019rga, Macpherson:2021lbr}, a comprehensive classification program has emerged to systematically chart the space of supersymmetric vacua with varying amounts of preserved supercharges and diverse internal geometries. Maximal supersymmetry cases, such as large $\mathcal{N} = (4,4)$ and $\mathcal{N} = (0,8)$, are now fully classified~\cite{DHoker:2008lup, Estes:2012vm, Bachas:2013vza, Macpherson:2018mif, Legramandi:2020txf, Dibitetto:2018ftj}, while sectors with reduced supersymmetry --including small $\mathcal{N} = (4,4)$ and $\mathcal{N} = (0,2)$-- have been partially explored through local classifications and explicit solution families~\cite{Dibitetto:2020bsh, Gauntlett:2007ph, Couzens:2017nnr, Couzens:2019iog, Couzens:2022agr, Conti:2024rwd, Lozano:2022ouq, Capuozzo:2024onf, Conti:2024rqy, Kim:2005ez, Ramirez:2025jut}. Additional progress has been made in the classification of solutions preserving small $\mathcal{N} = (0,4)$ supersymmetry~\cite{Couzens:2017way, Lozano:2019emq, Macpherson:2022sbs,Faedo:2020lyw,Faedo:2020nol, Lozano:2015bra,Lozano:2020bxo, Lozano:2022ouq}. Meanwhile, several isolated examples are known for intermediate supersymmetry, such as $\mathcal{N} = (0,7)$, $(4,2)$, $(3,3)$, and $(2,1)$, although a systematic understanding remains out of reach (see~\cite{Tong:2014yna, Lozano:2015cra, Lozano:2015bra, Kelekci:2016uqv, Couzens:2017way, Eberhardt:2017pty, Dibitetto:2017tve, Dibitetto:2017klx, Datta:2017ert, Couzens:2017nnr, Gaberdiel:2018rqv, Eberhardt:2018sce, Dibitetto:2018ftj, Dibitetto:2018iar, Eberhardt:2018ouy, Macpherson:2018mif, Lozano:2019emq, Lozano:2019jza, Lozano:2019zvg, Lozano:2019ywa, Passias:2019rga, Eberhardt:2019ywk, Couzens:2019iog, Couzens:2019mkh, Legramandi:2019xqd, Lozano:2020bxo, Faedo:2020nol, Dibitetto:2020bsh, Passias:2020ubv, Faedo:2020lyw, Legramandi:2020txf, Couzens:2021tnv, Couzens:2021veb, Macpherson:2021lbr, Macpherson:2022sbs, Conti:2024rqy, Conti:2024rwd} for a more exhaustive list).
Among all known sectors, the case of $\mathcal{N} = (0,6)$ supersymmetry stands out as both particularly rich and largely unexplored. Two distinct classes of superconformal algebras can arise in this context: a \emph{large} algebra, $\mathfrak{osp}(6|2)$, with an enhanced R-symmetry group $\mathrm{SO}(6)$, and a \emph{small} algebra, $\mathfrak{su}(1,1|3)$, where the R-symmetry is reduced to $\mathrm{U}(3)$. The large algebra signals a higher degree of symmetry and is typically associated with highly symmetric internal geometries. In contrast, the small algebra appears in more generic settings where the internal manifold only preserves a $\mathrm{U}(3)$ symmetry. Despite their theoretical appeal, explicit examples of $\mathrm{AdS}_3$ solutions realising either class of $\mathcal{N} = (0,6)$ supersymmetry had remained elusive until very recently. The construction in~\cite{Macpherson:2023cbl}\footnote{In that work, $\mathrm{AdS}_3$ solutions with $\mathcal{N} = (0,5)$ and $\mathfrak{osp}(5|2)$ superconformal symmetry were also identified.} marks a major step forward and opens the door to a wealth of new possibilities in the study of $\mathrm{AdS}_3/\mathrm{CFT}_2$ holography.
	
In this proceedings contribution, we focus on a newly uncovered class of $\mathrm{AdS}_3$ solutions in type IIA supergravity preserving $\mathcal{N} = (0,6)$ supersymmetry~\cite{Macpherson:2023cbl}, along with their $\mathcal{N} = (0,4)$ realisation in type IIB~\cite{Lozano:2024idt}. These backgrounds are expected to arise as near-horizon geometries of intersecting brane configurations, and they admit a natural holographic interpretation in terms of strongly coupled two-dimensional conformal field theories supported on defect or interface sectors~\cite{Lozano:2024idt}. Their high degree of supersymmetry, together with a non-trivial $\mathrm{SO}(6)$ R-symmetry, makes them an ideal setting for precision tests of holography in lower dimensions. In particular, they offer a controlled yet non-trivial laboratory to investigate the dynamics of $\mathcal{N} = (0,6)$ SCFTs.

This proceedings is organised as follows. In Section~\ref{sec:2}, we review the construction of the $\mathrm{AdS}_3$ solutions in massive type IIA supergravity presented in~\cite{Macpherson:2023cbl}, which preserve $\mathcal{N}=(0,6)$ supersymmetry and realise an $\mathfrak{osp}(6|2)$ superconformal algebra. We focus on the class with $\mathrm{SO}(6)$ isometry, described in terms of a $\mathbb{CP}^3$ manifold parametrised as a fibration of $T^{1,1}$ over an interval. We present the relevant supersymmetry conditions, the associated $G$-structure bilinears, and derive explicit expressions for the fluxes and warp factors in terms of a single cubic function. In Section~\ref{sec:3}, we analyse the type IIA $\mathcal{N}=(0,6)$ backgrounds in detail and present their type IIB realisation obtained via Abelian T-duality along the $\psi$-direction. The dual backgrounds preserve $\mathcal{N}=(0,4)$ supersymmetry and admit an explicit expression for the fluxes and geometry. Section~\ref{sec:4} is devoted to the massless case, where the ABJM/ABJ solution is recovered from a quadratic profile. Section~\ref{sec:5} discusses the field theory dual to the massive backgrounds. We propose a brane box construction in type IIB involving D3-, NS5-, $(1,k)$ 5$'$-, D7- and fractional D5-branes, and show that it realises a 2d quiver gauge theory with $\mathcal{N}=(0,3)$ supersymmetry. We derive the matter content, analyse anomaly cancellation, and discuss Seiberg-like dualities induced by large gauge transformations.

\section{AdS$_3$ solutions in massive IIA supergravity with $\mathfrak{osp}(6|2)$ superconformal symmetry}\label{sec:2}
In \cite{Macpherson:2023cbl}, two new AdS$_3$ solutions of type IIA supergravity were constructed, preserving SO(5) and SO(6) isometries, with the internal space realized as a squashed \( \mathbb{CP}^3 \). In this section, we briefly review their construction, focusing specifically on the SO(6)-invariant backgrounds in the \( T^{1,1} \) parametrization, which will be the main subject of our analysis.

In general, AdS$_3$ backgrounds preserving the full SO(2,2)$\times$SO($n$) isometry, for $n=5,6$, take the form of warped products involving AdS$_3$ and a seven-dimensional internal space $M_7$. The space $M_7$ admits an SO($n$) symmetry, identified with the R-symmetry of the \( \mathcal{N} = (0,n) \) superconformal algebra \( \mathfrak{osp}(n|2) \). In particular, the \( \mathcal{N} = (0,6) \) solutions with SO(6) symmetry resemble the well-known AdS$_4 \times \mathbb{CP}^3$ background, where \( \mathbb{CP}^3 \) naturally exhibits the required symmetry structure. A useful parametrization employed in \cite{Macpherson:2023cbl} expresses \( \mathbb{CP}^3 \) as an S$^2$ fibration over S$^4$, allowing a controlled squashing that breaks SO(6) $\to$ SO(5) by reducing the fiber size. This choice is particularly convenient, as the fundamental representation of SO(6) decomposes under SO(5) as \( \mathbf{6} \to \mathbf{1} \oplus \mathbf{5} \), which aligns with the structure of supersymmetry-preserving backgrounds.

The metric and flux ansatz take the form,
\begin{align}
	&ds^2 = e^{2A} ds^2(\text{AdS}_3) + e^{2k} dr^2 + ds^2(\widehat{\mathbb{CP}}^3),\quad\quad H = e^{3A} h_0\; \text{vol}(\text{AdS}_3) + H_3,\nonumber\\
	&F = f_{\pm} + e^{3A} \text{vol}(\text{AdS}_3) \star_7 \lambda(f_{\pm}), \label{eq:defCP3squahsed}\\[2mm]
	&ds^2(\widehat{\mathbb{CP}}^3) = \frac{1}{4}\left[e^{2C} \left(d\alpha^2 + \frac{1}{4} \sin^2\alpha (L_i)^2\right) + e^{2D} (Dy_i)^2 \right], \quad Dy_i = dy_i + \cos^2\left(\frac{\alpha}{2}\right) \epsilon_{ijk} y_j L_k,\nonumber
\end{align}
Here, we explicitly display the parametrization of the squashed \( \widehat{\mathbb{CP}}^3 \), where $y_i$ are coordinates on the unit two-sphere, $L_i$ are SU(2) left-invariant one-forms, and all warp factors \( (e^{2A}, e^{2C}, e^{2D}, e^{2k}, \Phi) \) depend solely on $r$.

To focus on the \( \mathcal{N} = (0,6) \) solutions, we set \( e^{2C} = e^{2D} \), yielding a round \( \mathbb{CP}^3 \) of unit radius. This space is a K\"ahler--Einstein manifold with an SO(6)-invariant K\"ahler form $J$ and with the following SO(6)-invariants
\begin{equation}
	J, \quad\quad J\wedge J, \quad\quad J\wedge J\wedge J = 6\, \text{vol}(\mathbb{CP}^3).
\end{equation}
This form also satisfies \( dJ = 0 \).  Thus, the ansatz for the fluxes then reads,
\begin{align}
	H_3 &= dB_2, \quad B_2 = b(r) J, \label{eq:magneticNSflux}\\
	f_+ &= \left[F_0 + c_1 J + c_2 J \wedge J + c_3 \frac{1}{3!} J \wedge J \wedge J\right] \wedge e^{B_2}.\nonumber
\end{align}
This SO(6)-invariant ansatz allows us to impose supersymmetry. For AdS$_3$ solutions in type II supergravity to preserve $\mathcal{N} = (0,1)$ supersymmetry, the following conditions must be satisfied \cite{Dibitetto:2018ftj, Passias:2019rga, Macpherson:2021lbr},
\begin{subequations}
	\label{BPS}
	\begin{align}
		d_{H_3}(e^{2A - \Phi}  \Psi_{\mp}) &= 0, \label{eq:SUSYI}\\[2mm]
		d_{H_3}(e^{3A - \Phi}  \Psi_{\pm}) \mp 2m e^{2A - \Phi}  \Psi_{\mp} &= \frac{e^{3A}}{8} \star_7 \lambda(f_{\pm}), \label{eq:SUSYII}\\[2mm]
		e^{A} (\Psi_{\mp}, f_{\pm})_7 &= \mp \frac{m}{2} e^{-\Phi}  \text{vol}(M_7), \label{eq:projection}
	\end{align}
\end{subequations}
where \( (\Psi_{\mp}, f_{\pm})_7 \) denotes the 7-form component of the wedge product \( \Psi_{\mp} \wedge \lambda(f_{\pm}) \). The bispinors \( \Psi_\pm \) are built from the internal seven dimensional spinors \( (\chi_1, \chi_2) \) as
\begin{equation}
	\Psi_+ + i \Psi_-= \frac{1}{8} \sum_{n=0}^7 \frac{1}{n!} \chi_2^{\dag} \gamma_{a_n \ldots a_1} \chi_1\, e^{a_1 \ldots a_n} . \label{eq:map}
\end{equation}
Here, $e^a$ denotes a vielbein on $M_7$. 

The  $\mathcal{N} = 6$ seven dimensional spinors  were constructed in \cite{Macpherson:2023cbl} in terms of two real functions of $r$, $\beta_1=\beta_1(r)$, $\beta_2=\beta_2(r)$,  and spinors on $\mathbb{CP}^3$, $\zeta_6^\mathcal{I}$ (for $\mathcal{I}=1,...,6$), as follows,
\begin{align}
	\chi_1^\mathcal{I}&=\cos\left(\frac{\beta_1+\beta_2}{2}\right)\; \zeta_6^\mathcal{I}+\sin\left(\frac{\beta_1+\beta_2}{2}\right)\; \hat{\zeta}_6^\mathcal{I},\nonumber\\ \chi_2^\mathcal{I}&=\cos\left(\frac{\beta_1-\beta_2}{2}\right)\; \zeta_6^\mathcal{I}+\sin\left(\frac{\beta_1-\beta_2}{2}\right)\; \hat{\zeta}_6^\mathcal{I}.\label{eq:7dimSpinors}
\end{align}
These internal six-dimensional spinors, \(\zeta_6^\mathcal{I}\), on \(\mathbb{CP}^3\) were derived in~\cite{Ramirez:2025jut,Conti:2025djz} starting from the Killing spinors on the seven-sphere, which admits a natural description as a \(U(1)\) fibration over \(\mathbb{CP}^3\). The Killing spinors on \(S^7\) satisfy the standard Killing spinor equation and transform in the \(\mathbf{8}\) of \(\mathrm{SO}(8)\), the isometry group of the round sphere. Among the two sets of spinors appearing in the full \(\mathcal{N}=8\) spectrum, only one realizes the branching \(\mathbf{8} \to \mathbf{6}_0 \oplus \mathbf{1}_{+1} \oplus \mathbf{1}_{-1}\) under the subgroup \(\mathrm{SO}(6) \times \mathrm{SO}(2)\). Upon reducing along the U(1) fiber, the \(\mathbf{6}_0\) components, being uncharged under the isometry, survive and descend to \(\mathbb{CP}^3\), while the \(\mathbf{1}_{\pm1}\) components, which are charged under \(\mathrm{SO}(2)\), are projected out. This leads to six spinors transforming in the \(\mathbf{6}\) of \(\mathrm{SO}(6)\), which define the internal spinors \(\zeta_6^\mathcal{I}\) and span the preserved supersymmetries of the background. These spinors, plugged into the expressions \eqref{eq:7dimSpinors}-\eqref{eq:map}, give rise to six pairs of bilinears,
\begin{align}
	\label{bilinears}
	\Psi_+^{(\mathcal{I})} &= \frac{1}{16}\, \text{Re}\left[e^{i\beta_2} e^{-i\mathcal{J}^{(\mathcal{I})}} - e^{i\beta_1} \Omega^{(\mathcal{I})} \wedge V\right],\nonumber \\
	\Psi_-^{(\mathcal{I})} &= \frac{1}{16}\, \text{Im}\left[e^{i\beta_2} e^{-i\mathcal{J}^{(\mathcal{I})}} \wedge V + e^{i\beta_1} \Omega^{(\mathcal{I})}\right],
\end{align}
which are spanned by the forms \((\mathcal{J}^{(\mathcal{I})}, \Omega^{(\mathcal{I})})\). All of these forms are charged under \(\mathrm{SO}(6)\) and satisfy the SU(3)-structure conditions on $M_7$,
\begin{align}
	\label{eq:SU(3)restrictionsonM7}
	\iota_V \mathcal{J}^{(\mathcal{I})} = \iota_V \Omega^{(\mathcal{I})} = 0, \quad \mathcal{J}^{(\mathcal{I})} \wedge \Omega^{(\mathcal{I})} = 0, \quad \mathcal{J}^{(\mathcal{I})}\wedge\mathcal{J}^{(\mathcal{I})}\wedge\mathcal{J}^{(\mathcal{I})} = \frac{3i}{4} \Omega^{(\mathcal{I})} \wedge \bar{\Omega}^{(\mathcal{I})}.
\end{align}
In addition to the charged SU(3)-structures, one can also construct bilinears from the singlet spinors in the \(\mathbf{1}_{\pm1}\) representations, which give rise to the forms \((J, e^\tau, \Omega_3^{(a)})\) on \(S^7\). These forms are invariant under \(\mathrm{SO}(6)\) by construction, but only \(J\) is uncharged under the \(U(1)\) isometry and therefore survives the reduction to \(\mathbb{CP}^3\).  As a result, \(J\) appears explicitly in the flux ansatz \eqref{eq:magneticNSflux} and plays the role of the canonical K\"{a}hler form on \(\mathbb{CP}^3\), while the six charged SU(3)-structures are responsible for capturing the full \(\mathcal{N}=6\) supersymmetry preserved by the background.

The bilinears in equation~\eqref{bilinears} solve the BPS conditions \eqref{BPS} for the background fields,
\begin{align}
	\frac{ds^2}{2\pi} &= \frac{|hu|}{\sqrt{\Delta_1}} ds^2(\text{AdS}_3) + \frac{\sqrt{\Delta_1}}{|u|} \left[ \frac{2}{|h''|} ds^2(\mathbb{CP}^3) + \frac{1}{4|h|} dr^2 \right], \nonumber\\[2mm]
	e^{-\Phi} &= \frac{ \sqrt{|u|} |h''|^{3/2} \sqrt{\Delta_2} }{2\sqrt{\pi} \Delta_1^{1/4}}, \quad \Delta_1 = 2hh''u^2 - (uh' - hu')^2, \quad \Delta_2 = 1 + \frac{2h'u'}{u h''}, \nonumber\\[2mm]
	B_2 &= 4\pi \left[-(r - k) + \frac{uh' - hu'}{u h''} + \frac{u'}{2h''} \left( \frac{h}{u} + \frac{hh'' - 2(h')^2}{2h'u' + u h''} \right) \right] J. \label{eq:summerystart}
\end{align}
Here, $h(r)$ and $u(r)$ are functions of $r$, and $k$ is a constant. The RR sector is given by,
\begin{align}
	\label{RRgeometry}
	F_0 &= -\frac{1}{2\pi} h''', \quad
	F_2 = B_2 F_0 + 2(h'' - (r - k) h''') J, \\
	F_4 &= \pi d\left(h' + \frac{hh''u(uh' + hu')}{\Delta_1}\right) \wedge \text{vol}(\text{AdS}_3) + B_2 \wedge F_2 - \frac{1}{2} B_2 \wedge B_2 F_0 \nonumber\\
	&\quad -4\pi (2h' + (r - k)(-2h'' + (r - k) h''')) J \wedge J.
\end{align}

Solutions in this class are locally determined by two differential equations. First, supersymmetry requires
\begin{equation}
	\label{eq:bpscondtion}
	u = \text{const},
\end{equation}
which must hold globally. Without loss of generality, we will fix \( u = 1 \) for the remainder of this work. Second, the Bianchi identity for $F_0$ demands that in regular regions of the internal space,
\begin{equation}
	\label{eq:summeryend}
	h'''' = 0,
\end{equation}
which integrates to $h = c_0 + c_1 r + \tfrac{1}{2} c_2 r^2 + \tfrac{1}{3!} c_3 r^3$, with $c_i$ constants. Globally, \eqref{eq:summeryend} may receive $\delta$-function source terms, as we will explore later.

As noted earlier, the bilinears in \eqref{bilinears} solve the supersymmetry constraints \eqref{BPS} for the background fields in \eqref{eq:summerystart} and \eqref{RRgeometry}, thereby ensuring the preservation of \( \mathcal{N} = (0,1) \) supersymmetry. By considering the full set of spinor pairs, one generates five additional independent bispinors that also satisfy \eqref{BPS}, resulting in a total of \( \mathcal{N} = (0,6) \) supersymmetry. 

In the next section, we will consider a T-duality along the \(\psi\) direction, which corresponds to a natural U(1) isometry in the parametrisation of \(\mathbb{CP}^3\) as a fibration of \(T^{1,1}\) over an interval. In order to determine the amount of supersymmetry preserved by the dual background, we focus on the spinors that are invariant under translations along \(\partial_\psi\). As shown in~\cite{Ramirez:2025jut}, this isometry acts non-trivially only on two of the six internal spinors. Consequently, the T-dual background preserves the four spinors that are neutral under this action, leading to four corresponding bispinors that survive the duality. This implies that the resulting geometry preserves \(\mathcal{N} = (0,4)\) supersymmetry.

\section{The type IIA $\mathcal{N}=(0,6)$ class and its type IIB realisation}\label{sec:3}

This section is devoted to the analysis of the massive type IIA $\mathcal{N} = (0,6)$ solutions constructed in~\cite{Macpherson:2023cbl}, as well as their $\mathcal{N} = (0,4)$ type IIB counterparts obtained via T-duality in~\cite{Lozano:2024idt}. These two classes of solutions form the backbone of the constructions developed in the remainder of this work.

The bilinears defined in equation~\eqref{bilinears} solve the supersymmetry conditions~\eqref{BPS} with $u=1$ for the following NS-NS fields,
\begin{align}
	\label{eq:metric}
	\frac{ds^2}{2\pi} &= \frac{h}{\Delta}\,ds^2(\text{AdS}_3) + \Delta \left[ \frac{1}{4h}dr^2 + \frac{2}{h''}ds^2(\mathbb{CP}^3) \right], \qquad
	e^{-2\Phi} = \frac{(h'')^3}{4\pi\Delta}, \nonumber\\[2mm]
	B_2 &= 4\pi \left( -(r - l) + \frac{h'}{h''} \right) J, \qquad
	\Delta = \sqrt{2hh'' - (h')^2}.
\end{align}

The corresponding RR field strengths are,
\begin{align}
	F_0 &= -\frac{h'''}{2\pi}, \qquad 
	F_2 = B_2 F_0 + 2(h'' - (r - l)h''') J, \nonumber\\[2mm]
	F_4 &= -\pi d\left[h' + \frac{hh'h''}{\Delta^2}\right] \wedge \text{vol}(\text{AdS}_3) 
	+ B_2 \wedge \left(F_2 - \tfrac{1}{2} B_2 F_0\right) \nonumber\\[2mm]
	&\quad - 4\pi\left(2h' + (r - l)\left(-2h'' + (r - l)h'''\right)\right) J \wedge J. \label{eq:RRgeometry}
\end{align}

The internal manifold $\mathbb{CP}^3$ is written as a foliation of $T^{1,1}$ over an interval, with the metric and associated forms given by,
\begin{align}
	\label{eq:paramCP3}
	ds^2(\mathbb{CP}^3) &= d\xi^2 + \frac{1}{4} c_\xi^2\, ds^2(\text{S}_1^2) + \frac{1}{4} s_\xi^2\, ds^2(\text{S}_2^2) + \frac{1}{4} s_\xi^2 c_\xi^2\, (d\psi + \eta_1 - \eta_2)^2, \\[2mm]
	d\eta_i &= -\text{vol}(\text{S}^2_i), \qquad J= \frac{1}{4}  s_\xi^2\, \text{vol}(\text{S}_2^2) + \frac{1}{4} c_\xi^2\, \text{vol}(\text{S}_1^2) 
	+ \frac{1}{2} s_\xi c_\xi\, d\xi \wedge (d\psi + \eta_1 - \eta_2),\nonumber 
\end{align}
where both 2-spheres have unit radius.

In regular regions, the Bianchi identity for $F_0$ implies that it must be constant. This yields the differential equation,
\begin{equation}\label{eq:definingPDE}
	h''' = -2\pi F_0,
\end{equation}
whose general solution is a cubic polynomial in $r$. The qualitative behavior of the solution, including its physical interpretation and domain, depends on how the coefficients of $h$ are chosen. Assuming $F_0 \neq 0$ and that the $r$-interval is bounded at one end (taken without loss of generality to be $r = 0$), three distinct physical singularities were identified in \cite{Macpherson:2023cbl}: a D8/O8 system, an O2-plane sitting at the tip of a G$_2$ cone with $\mathbb{CP}^3$ base and a $d = 3$ KK-monopole singularity arising from the reduction of the 8d hyper-K\"{a}hler manifolds constructed in \cite{Gauntlett:1997pk}.

As emphasized in \cite{Macpherson:2023cbl}, $F_0$ can also be piecewise constant. If it jumps across a locus $r = r_0$, the Bianchi identity implies,
\begin{equation}
	dF_0 = \Delta F_0\, \delta(r - r_0),
\end{equation}
signaling the presence of a stack of D8-branes with charge $2\pi \Delta F_0$. Arbitrary numbers of D8-branes may be introduced along the $r$-interval, provided they lie at integer positions and are accompanied by the appropriate large gauge transformations of $B_2$.

A more tractable realization of these geometries is obtained via T-duality to type IIB, where the dual CFT --particularly in the ABJM/ABJ context-- is better understood. In \cite{Lozano:2024idt}, the type IIB background was obtained by performing an Abelian T-duality along the U(1) isometry of $T^{1,1}$, parametrized by $\psi$ in \eqref{eq:paramCP3}. The resulting NS-NS fields are,
\begin{align}
	\begin{split}\label{IIBmetric}
		\frac{ds^2}{2\pi}=&\frac{h}{\Delta}\,ds^2(\text{AdS}_3)+\frac{2{\Delta}}{h''}\left(d\xi^2+\frac14 c_\xi^2 ds^2(\text{S}_1^2)+\frac14s_\xi^2 ds^2(\text{S}_2^2)\right)+\frac{{\Delta}}{4h}\, dr^2\\
		&+\frac{2h''}{\Delta}\,\left[\left(\frac{h'}{h''}-r\right)d\xi-\frac{d\psi}{2\pi s_\xi c_\xi}\right]^2\, ,\qquad e^{2\Phi}=\frac{4}{s_\xi^2c_\xi^2\,(h'')^2}\,,\\
		B_2=&\pi \left(r-\frac{h'}{h''} \right)\,\bigl(c_\xi^2\,\text{vol}(\text{S}_1^2)-s_\xi^2\,\text{vol}(\text{S}_2^2)\bigr)+\left(\eta_1+\eta_2\right)\wedge d\psi\,,
	\end{split}
\end{align}
and the corresponding RR field strengths are,
\begin{align}
	\label{RRIIBfluxes}
	& F_1=s_\xi c_\xi\,\left(h''-r h''' \right)d\xi-\frac{h'''}{2\pi}\,d\psi\,,\nn\\
	& F_3=\pi s_\xi c_\xi\,\left(\frac{h'(h''+rh''')-r(h'')^2}{h''}\,d\xi-\frac{(h'')^2-h'h'''}{h''}\,\frac{d\psi}{2\pi s_\xi c_\xi}\right)\wedge\bigl(c_\xi^2\,\text{vol}(\text{S}_1^2)-s_\xi^2\,\text{vol}(\text{S}_2^2)\bigr)\,,\nn\\
	&F_5=2\pi^2 s_\xi c_\xi \left(\left(rh'-2h-\frac{hh'(h'-rh'')}{\Delta^2} \right)'d\xi+\frac12\,\left( 3h'+\frac{(h')^3}{\Delta^2} \right)'\frac{d\psi}{2\pi s_\xi c_\xi}\right)\wedge\text{vol}(\text{AdS}_3)\wedge dr\nn\\
&\qquad+\pi^2s_\xi^3c_\xi^3\left(\frac{r(h')^2h'''-3h''\Delta^2}{(h'')^2}\,d\xi+\frac{(h')^2h'''}{(h'')^2}\,\frac{d\psi}{2\pi s_\xi c_\xi}\right)\wedge \text{vol}(\text{S}_1^2)\wedge \text{vol}(\text{S}_2^2)\,,
\end{align}
The defining function $h(r)$ continues to satisfy $h''' = -2\pi F_0$, with $F_0$ now interpreted as the D7-brane charge. Discontinuities in $F_0$ correspond to smeared D7-brane sources along the $\psi$-direction.

As discussed in the previous section, the internal spinors \(\zeta_6^\mathcal{I}\) on \(\mathbb{CP}^3\) include four components that are uncharged under the isometry generated by \(\partial_\psi\), where \(\psi\) parametrises the U(1) fiber in the \(T^{1,1}\) realisation of \(\mathbb{CP}^3\). Since the T-duality to type IIB is performed along this \(\psi\) direction, only the spinors invariant under this isometry contribute to the supersymmetry of the dual background. Consequently, four of the six bispinors \(\Psi_\pm^{(\mathcal{I})}\) survive the dualisation, leading to a type IIB solution that preserves \(\mathcal{N} = (0,4)\) supersymmetry.

To conclude this section, we present the expression for the holographic central charge. Following the results in \cite{Klebanov:2007ws,Macpherson:2014eza,Bea:2015fja}, which apply to the class of solutions described by \eqref{eq:metric}, the central charge takes the simple form
\begin{equation}\label{centralcharge}
	c_{\text{hol}} = \frac{1}{2} \int dr \left(2h\,h'' - (h')^2\right).
\end{equation}

Alternatively, and as we will see later, the central charge can be related to the Page charges of the background, computed from the Page fluxes
\begin{align}
	\hat f_2 &=  2\left(h'' - (r - l) h'''\right) J, \nonumber\\[2mm]
	\hat f_4 &= -4\pi \left(2h' + (r - l)\left((r - l) h''' - 2h''\right)\right) J \wedge J, \nonumber\\[2mm]
	\hat f_6 &= \frac{16\pi^2}{3} \left(6h - (r - l)\left(6h' + (r - l)\left((r - l) h''' - 3h''\right)\right)\right) J \wedge J \wedge J. \label{eq:pagefluxes}
\end{align}
These fluxes will play a central role in the following sections, where we compute the quantised Page charges of D-branes and establish their relation to the holographic central charge.

\section{Massless case in type IIA and IIB: ABJ(M)}\label{sec:4}
As shown in \cite{Macpherson:2023cbl}  in the massless limit the ABJM/ABJ solution  \cite{Aharony:2008ug,Aharony:2008gk} is recovered. To see this we can parametrise
\begin{equation}\label{eq:AdS4h}
	h(r)=Q_2-Q_4r+\frac12 Q_6 r^2,
\end{equation}
where $Q_{2,4,6}$ are constants whose significance will become clear shortly. Then one can easily check that the change of variables
\begin{equation}\label{sinh}
	\sinh{\mu}=\frac{4\pi}{L^2Q_6}(Q_6 \,r -Q_4), \qquad \text{for}\qquad L^2Q_6=4\pi\sqrt{2Q_2Q_6-Q_4^2},
\end{equation}
gives rise to the ABJM/ABJ metric and dilaton
\begin{eqnarray}
	&&ds^2=L^2\,\Bigl(\frac14 ds^2(\text{AdS}_4)+ds^2(\mathbb{CP}^3)\Bigr),\quad e^{-2\Phi}=Q_6^{2}L^{-2},\quad B_2=-4\pi\frac{Q_4}{Q_6}J,  \label{ABJMdilaton}
\end{eqnarray} 
where $L$ is the radius of $\mathbb{CP}^3$ and the metric for AdS$_4$ is now parametrised as,
\begin{equation}
ds^2(\text{AdS}_4)=d\mu^2+\cosh^2{\mu} \,ds^2(\text{AdS}_3)
\end{equation} 
Therefore there is just one local solution when $F_0=0$ and it is an AdS$_4$ vacuum preserving twice the supersymmetries of generic solutions within this class. This is actually the only regular solution also.

Before moving on to the T-dual realisation, it is useful to interpret the parameters \( Q_{2,4,6} \) introduced in the massless profile~\eqref{eq:AdS4h} in terms of the underlying brane content. The identification of these constants as quantised charges follows from integrating the Page fluxes~\eqref{eq:pagefluxes} over the relevant cycles of \(\mathbb{CP}^3\),
\begin{equation}
	\frac{1}{2\pi} \int_{\mathbb{CP}^1} \hat f_2 = Q_6, \quad
	\frac{1}{(2\pi)^3} \int_{\mathbb{CP}^2} \hat f_4 = Q_4, \quad
	\frac{1}{(2\pi)^5} \int_{\mathbb{CP}^3} \hat f_6 = Q_2. \label{QsAdS4}
\end{equation}
This identification allows us to associate \( Q_p \) with the Page charges of D\(p\)-branes for \( p = 2,4,6 \), respectively. In the AdS\(_4\) background under consideration, the NS-NS 3-form flux vanishes, \( H_3 = 0 \), and one has the relation \( \hat f_4 = -f_2 \wedge B_2 \), indicating that NS5 and D4-branes are not physical sources. Instead, their fluxes effectively cancel, consistent with the mechanism described in~\cite{Aharony:2008gk}.

In the type IIB case, we also take the same $h$ function as \eqref{eq:AdS4h} and the same change of variables \eqref{sinh} thus we obtain,
\begin{equation}
	\begin{split}
		&\frac{ds^2}{2\pi}=\frac{L^2}{8\pi }\,\left(ds^2(\text{AdS}_4)+4d\xi^2+c_\xi^2 ds^2(\text{S}_1^2)+s_\xi^2 ds^2(\text{S}_2^2)\right)+\frac{8\pi Q_4^2}{L^2Q_6^2}\,\bigl(d\xi+\frac{Q_6}{2\pi Q_4s_\xi c_\xi}\,d\psi\bigr)^2\\
		& e^{2\Phi}=\frac{4}{Q_6^2\,s_\xi^2 c_\xi^2}\,,\qquad B_2=\frac{\pi Q_4}{Q_6} \,\bigl[c_\xi^2\,\text{vol}(\text{S}_1^2)-s_\xi^2\,\text{vol}(\text{S}_2^2)\bigr]+\left(\eta_1+\eta_2\right)\wedge d\psi\,.\\
	\end{split}
\end{equation}
This is equivalent to taking the T-dual of the AdS$_4\times \mathbb{CP}^3$ solution. In this limit supersymmetry is enhanced and the background \eqref{IIBmetric} takes the form of an $\mathcal{N}=4$ solution with topology AdS$_4\times \text{S}^2\times \text{S}^2\times \Sigma_2$, which were classified in \cite{DHoker:2007zhm}. From this expression we can extract the 2d metric over the Riemann surface $\Sigma_2$, parametrised by the coordinates $(\xi, \psi)$. We observe that this surface is an annulus.

The RR fluxes read
\begin{equation}
	\begin{split}\label{RRABJMlimit}
		& F_1=Q_6\,s_\xi c_\xi\, d\xi\,,\qquad F_3=-\frac{Q_6}{2}\,\left(\frac{2\pi Q_4}{Q_6}s_\xi c_\xi\,d\xi+d\psi   \right)\wedge\left(c_\xi^2\,\text{vol}(\text{S}_1^2)-s_\xi^2\,\text{vol}(\text{S}_2^2)\right)\,,\\
		&F_5=\frac{3L^2Q_6}{8}\,\text{vol}(\text{AdS}_4)\wedge\left(\frac{2\pi Q_4}{Q_6}s_\xi c_\xi d\xi+d\psi\right)+\frac{3\pi L^2}{4}s_\xi^3c_\xi^3\,d\xi\wedge \text{vol}(\text{S}_1^2)\wedge \text{vol}(\text{S}_2^2).
	\end{split}
\end{equation}

\begin{figure}[t]
	\centering
	\includegraphics[scale=0.7]{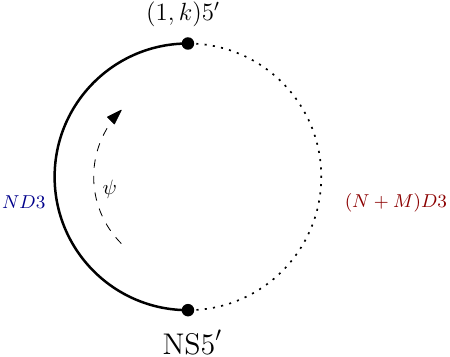}
	\caption{Type IIB brane configuration for the ABJM/ABJ theory. D3-branes wrap the $\psi$-circle and intersect an NS5$'$-brane and a $(1,k)$ 5$'$-brane, both extended along $(x^0, x^1, r)$.} \label{fig:ABJM}
\end{figure}

Table \ref{tableABJM} summarises the brane set-up associated to the ABJM/ABJ solution in its Type IIB realisation \cite{Aharony:2008ug}. Figure \ref{fig:ABJM} 
depicts the configuration: the D3-branes stretch along the $\psi$-circle, intersecting at $\psi = \pi$ a bound state consisting of a  NS5$'$ brane and D5$'$ branes, combined into a $(1,k)$ 5$'$-brane. This bound state is extended along $(x^0,x^1,r)$ and rotated by the same angle in the $[3,7]_\theta$, $[4,8]_\theta$, and $[5,9]_\theta$ planes relative to the NS5$'$ branes located at $\psi=0$ and $\psi=2\pi$, with $\tan\theta = k$. On top of this there are $M$ fractional branes stretched just along one segment of the circle, which as mentioned above are not actual branes since they are unstable. The brane system preserves $\mathcal{N}=3$ supersymmetry in 3d, and this is enhanced to $\mathcal{N}=6$ in the IR \cite{Aharony:2008ug}.
\begin{table}[t]
	\renewcommand{\arraystretch}{1}
	\begin{center}
		\scalebox{1}[1]{
			\begin{tabular}{c| c cc  c c  c  c c c c}
				branes & $x^0$ & $x^1$ & $r$ & $x^3$ & $x^4$ & $x^5$ & $\psi$ & $x^7$ & $x^8$ & $x^9$ \\
				\hline \hline
				$N\,\mrm{D}3$ & $\times$ & $\times$ & $\times$ & $-$ & $-$ & $-$ & $\times$ & $-$ & $-$ & $-$ \\
				$\mrm{NS}5'$ & $\times$ & $\times$ & $\times$ & $\times$ & $\times$ & $\times$ & $-$ & $-$ & $-$ & $-$ \\
				$(1,k) 5'$ & $\times$ & $\times$ & $\times$ & $\cos{\theta}$ & $\cos{\theta}$ & $\cos{\theta}$ & $-$ & $\sin{\theta}$ & $\sin{\theta}$ & $\sin{\theta}$ \\
			\end{tabular}
		}
		\caption{Brane configuration underlying the ABJM construction. The directions $(x^0, x^1, r)$ define the three-dimensional gauge theory. At $\psi = \pi$, a bound state formed by an NS5$'$ brane and D5$'$ branes is present, rotated by the same angle in the $(x^3, x^7)$, $(x^4, x^8)$, and $(x^5, x^9)$ planes relative to the NS5$'$ branes located at $\psi = 0$ and $\psi = 2\pi$. This setup preserves six Poincar\'e supercharges, corresponding to $\mathcal{N} = 3$ supersymmetry in three dimensions.
		} \label{tableABJM}
	\end{center}
\end{table}

\vspace{10 pt}
We begin by examining the massless limit in detail, both as a warm-up and because it represents the foundational conformal field theory. In this limit, large gauge transformations leave the theory invariant but induce shifts in the quantised charges\footnote{As shown in \cite{Lozano:2024idt}, this follows from the continuity of the solution.},
\begin{eqnarray}
		Q_2\rightarrow Q_2-Q_4+\frac12 Q_6, \qquad Q_4\rightarrow Q_4-Q_6. \label{Pagetransmassless}
	\end{eqnarray}
	These shifts must correspond to symmetries of the theory. As demonstrated in \cite{Aharony:2009fc}, they generate Seiberg dualities, linking the IR dynamics of 3d $\mathcal{N}=3$ Chern-Simons-matter theories. As we mentioned, the quantised charges differ from the brane numbers due to the topology of $\mathbb{CP}^3$. Specifically, the fractional flux needed to cancel the Freed-Witten anomaly on the non-spin $\mathbb{CP}^2$ and curvature corrections from wrapped branes \cite{Freed:1999vc,Bergman:2009zh} alter the charge assignments. This leads to the following relations \cite{Aharony:2009fc},
	\begin{eqnarray}
		Q_2= N+\frac{k}{12}, \qquad Q_4=M-\frac{k}{2}, \qquad Q_6= k. \label{change}
	\end{eqnarray}
	Under the transformation \eqref{Pagetransmassless}, the field theory parameters shift as
	\begin{equation}\label{Seiberg}
		N\rightarrow N+k-M, \qquad M\rightarrow M-k,
	\end{equation}
	mapping the theory with gauge group U$(N+M)_k \times$U$(N)_{-k}$ to U$(N)_k \times$U$(N-M+k)_{-k}$, and so on --realizing the Seiberg dualities from \cite{Aharony:2008gk,Aharony:2009fc}.
	
In turn, the central charge per unit $r$-interval becomes
	\begin{equation}\label{centralcharge2d}
		c_{hol} = 
		Nk - \frac12 M(M - k) - \frac{1}{24}k^2,
	\end{equation}
	which is invariant under \eqref{Seiberg}. To recover the correct scaling in 3d, we multiply by the AdS$_4$ radius squared --divided by $2\pi$ to match 2d/3d conventions--
	\begin{equation}
		c_{hol}^{(3d)} =
		= \frac{1}{k}\left(2Nk - M(M-k) - \frac{1}{12}k^2\right)^{3/2}. \label{cholmassless}
	\end{equation}
	This gives the free energy of the 3d CFT in the supergravity approximation, including $1/N$ corrections \cite{Bergman:2009zh,Bergman:2013qoa}.
	Alternatively, using the Maxwell D2-brane charge, one finds
	\begin{equation}
		c_{hol}^{(3d)} = 2\sqrt{2k}(Q_2^M)^{3/2} = 2^{3/2} k^2 \hat{\lambda}^{3/2},
	\end{equation}
	with $\hat{\lambda} = Q_2^M / k$ the shifted 't Hooft coupling \cite{Drukker:2010nc}. This matches the strong coupling expansion of the free energy computed via matrix model techniques on S$^3$ \cite{Drukker:2010nc,Herzog:2010hf,Fuji:2011km,Marino:2011eh}.

\section{The dual field theory, massive case}\label{sec:5}

Following \cite{Aharony:2008ug,Aharony:2008gk}, we propose a Type IIB brane configuration that realises a two-dimensional theory with $\mathcal{N}=(0,3)$ supersymmetry, expected to enhance to $\mathcal{N}=(0,6)$ in the infrared. This mirrors the supersymmetry enhancement observed in ABJ(M) theories, though preserving only half the original supercharges.

The setup involves D3-branes stretched along the $(r, \psi)$ directions, bounded by NS5 and NS5$'$-branes. Additionally, the configuration includes rotated $(1,k)$ 5$'$-branes, D7 flavour branes, and fractional D5-branes. The full brane configuration is summarized in Table~\ref{HWbranesetupIIB}.
\begin{table}[t]
	\renewcommand{\arraystretch}{1}
	\begin{center}
		\scalebox{1}[1]{
			\begin{tabular}{c| c cc  c c  c  c c c c}
				Branes & $x^0$ & $x^1$ & $r$ & $x^3$ & $x^4$ & $x^5$ & $\psi$ & $x^7$ & $x^8$ & $x^9$ \\
				\hline \hline
				D3 & $\times$ & $\times$ & $\times$ & $-$ & $-$ & $-$ & $\times$ & $-$ & $-$ & $-$ \\
				NS5$'$ & $\times$ & $\times$ & $\times$ & $\times$ & $\times$ & $\times$ & $-$ & $-$ & $-$ & $-$ \\
				$(1,k)$ 5$'$ & $\times$ & $\times$ & $\times$ & $\cos{\theta}$ & $\cos{\theta}$ & $\cos{\theta}$ & $-$ & $\sin{\theta}$ & $\sin{\theta}$ & $\sin{\theta}$ \\
				D7 & $\times$ & $\times$ & $-$ & $\times$ & $\times$ & $\times$ & $-$ & $\times$ & $\times$ & $\times$ \\
				NS5 & $\times$ & $\times$ & $-$ & $-$ & $-$ & $-$ & $\times$ & $\times$ & $\times$ & $\times$ \\
			\end{tabular}
		}
		\caption{Type IIB brane setup realizing a 2d theory with $\mathcal{N}=(0,3)$ supersymmetry. D3-branes stretch along $(r, \psi)$ within a brane box bounded by NS5 and 5$'$-branes.} \label{HWbranesetupIIB}
	\end{center}
\end{table}
T-duality reduces the supersymmetry from $\mathcal{N}=(0,6)$ to $\mathcal{N}=(0,4)$, and the subsequent rotation of the 5$'$-branes further breaks it to $\mathcal{N}=(0,3)$, provided the rotation is symmetric along the $[3,7]$, $[4,8]$, and $[5,9]$ directions. The addition of D7-branes preserves the remaining supercharges but introduces brane creation effects that modify the number of D5$'$-branes across the $r$-direction.

\begin{figure}[t]
	\centering
	\includegraphics[scale=0.6]{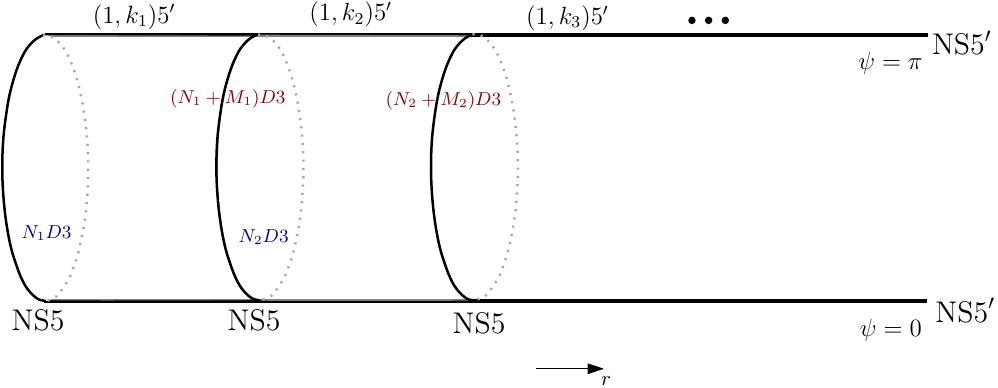}
	\caption{Brane box configuration: $N_l$ D3-branes extend from $\psi = 0$ to $\pi$, and $N_l + M_l$ from $\pi$ to $2\pi$.} \label{branediagram}
\end{figure}

\begin{figure}[t]
	\centering
	\includegraphics[scale=0.6]{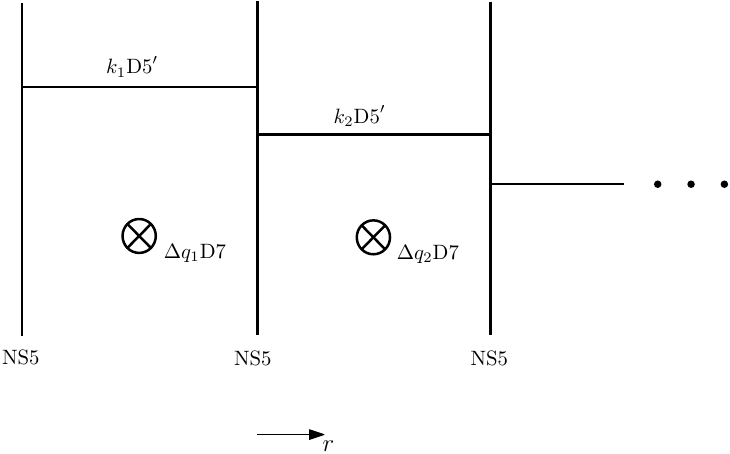}
	\caption{D5$'$-NS5-D7 subsystem illustrating brane creation due to the presence of D7-branes.} \label{D5-D7}
\end{figure}

This brane system defines a brane box model, as illustrated in Figure~\ref{branediagram}. In each segment, $N_l$ D3-branes are suspended between NS5-branes along $r$, and between an NS5$'$ brane at $\psi = 0$ and a $(1,k_l)$ 5$'$-brane at $\psi = \pi$. Between $\psi = \pi$ and $\psi = 2\pi$, the number increases to $N_l + M_l$ due to the creation of fractional D5-branes.
As one moves along $r$, the number of D5$'$-branes $k_l$ changes due to a brane creation mechanism triggered by the D7-branes. This leads to a variation in the rotation angle $\theta_l$, governed by $\tan \theta_l = \Delta k_l$. The local structure of this subsystem, depicted in Figure~\ref{D5-D7}, involves $k_l$ D5$'$-branes stretched between NS5-branes and $\Delta q_l$ D7-branes located within each interval $r \in [l, l+1]$.

This Type IIB brane box configuration realises a 2d field theory on the D3-branes, with $\mathcal{N}=(0,3)$ supersymmetry and a non-trivial dependence on the distribution of D5$'$ and D7-branes. 

Since little is currently known about two-dimensional $\mathcal{N}=(0,3)$ supersymmetric theories, we adopt a phenomenological approach in this subsection, hoping it may inspire further detailed study. As a starting point, we review the $\mathcal{N}=(0,4)$ brane box models introduced in \cite{Hanany:2018hlz}, which form the foundation of our construction. Our setup extends these models by incorporating a relative rotation between the D5$'$- and NS5$'$-branes.

The original brane box model consists of a D3-NS5-NS5$'$-D5-D5$'$ intersection with branes oriented as shown in Table~\ref{HO}. This system preserves $\mathcal{N}=(0,4)$ supersymmetry.
\begin{table}[t]
	\renewcommand{\arraystretch}{1}
	\begin{center}
		\scalebox{1}[1]{
			\begin{tabular}{c| c cc  c c  c  c c c c}
				Brane & $x^0$ & $x^1$ & $r$ & $x^3$ & $x^4$ & $x^5$ & $\psi$ & $x^7$ & $x^8$ & $x^9$ \\
				\hline \hline
				D3        & $\times$ & $\times$ & $\times$ & $-$ & $-$ & $-$ & $\times$ & $-$ & $-$ & $-$ \\
				NS5$'$    & $\times$ & $\times$ & $\times$ & $\times$ & $\times$ & $\times$ & $-$ & $-$ & $-$ & $-$ \\
				D5$'$     & $\times$ & $\times$ & $\times$ & $-$ & $-$ & $-$ & $-$ & $\times$ & $\times$ & $\times$ \\
				NS5       & $\times$ & $\times$ & $-$ & $-$ & $-$ & $-$ & $\times$ & $\times$ & $\times$ & $\times$ \\
				D5        & $\times$ & $\times$ & $-$ & $\times$ & $\times$ & $\times$ & $\times$ & $-$ & $-$ & $-$ \\
			\end{tabular}
		}
		\caption{Brane configuration of the $\mathcal{N}=(0,4)$ D3-brane box model from \cite{Hanany:2018hlz}. The field theory lives along $(x^0,x^1)$, with $r$ and $\psi$ representing the spatial directions of the brane box.} \label{HO}
	\end{center}
\end{table}
In this setup, the global symmetry group $\text{SO}(3)_{345} \times \text{SO}(3)_{789}$ is identified with the $\text{SO}(4)_R$ R-symmetry of the $\mathcal{N}=(0,4)$ theory. In our extension, the NS5$'$ brane and the D5$'$ branes are placed at the same position along $\psi$, forming a bound state that is rotated by a common angle in the $[3,7]$, $[4,8]$, and $[5,9]$ planes relative to the NS5$'$ branes located at $\psi=0$. This rotation induces masses for some scalar modes in the spectrum, as discussed in \cite{Kitao:1998mf,Bergman:1999na}.

\begin{figure}[t]
	\centering
	\includegraphics[scale=0.75]{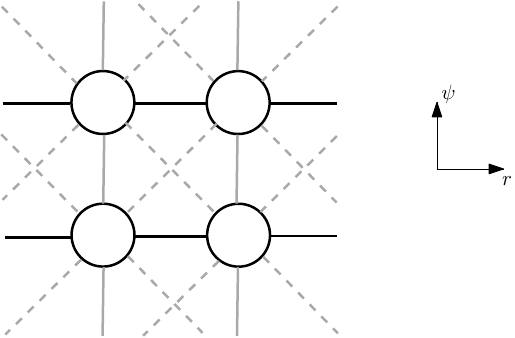}
	\caption{Quiver diagram corresponding to D3-branes suspended between NS5- and NS5$'$-branes along orthogonal directions. Circles indicate $(0,4)$ vector multiplets; black, gray, and dashed lines represent twisted hypermultiplets, hypermultiplets, and Fermi multiplets, respectively.} \label{HOquiver}
\end{figure}

The low-energy spectrum of the D3-brane box model \cite{Hanany:2018hlz} without rotation contains four types of $\mathcal{N}=(0,4)$ multiplets, as detailed in Table~\ref{hypers}. These arise from the quantisation of open strings ending on D3-branes, with different boundary conditions set by their relative positions across NS5 and NS5$'$ branes. Additional contributions from D5$'$ and D7-branes modify this matter content, as also shown in the lower rows of Table~\ref{hypers}.
\begin{table}[t]
	\renewcommand{\arraystretch}{1.2}
	\begin{center}
		\scalebox{1}[1]{
		\begin{tabular}{c| c| c| c  }
			String & Multiplet & Interval & Symbol \\
			\hline \hline
			$\mrm{D}3$-$\mrm{D}3$ & (0,4) twisted hyper bifundamental & separated by a NS5 &  --- \\
			$\mrm{D}3$-$\mrm{D}3$ & (0,4) vector & same stack & $\bigcirc$  \\
			$\mrm{D}3$-$\mrm{D}3$ &Two (0,4) hyper bifundamental & separated by a NS5' & {\color{lightgray} ---}  \\
			$\mrm{D}3$-$\mrm{D}3$ &Two (0,2) Fermi=(0,4) Fermi & separated by a NS5 and NS5' & - - -   \\
			\hline \hline\hline 
			$\mrm{D}3$-$\mrm{D}5'$ & (0,4) hyper bifundamental & same interval &  {\color{lightgray} ---}  \\
			$\mrm{D}3$-$\mrm{D}5'$ & (0,2) Fermi & adjacent interval & {\color{lightgray} - - -}  \\
			$\mrm{D}3$-$\mrm{D}7$ & (0,2) Fermi & same interval & {\color{lightgray} - - -}  \\
		\end{tabular}
		}
		\caption{Matter content in the $\mathcal{N}=(0,4)$ D3-brane box model. The top block lists multiplets from D3-D3 strings \cite{Hanany:2018hlz}; the bottom block includes contributions from D5$'$ and D7 flavour branes.} \label{hypers}
	\end{center}
\end{table}
Figure~\ref{HOquiver} illustrates the general structure of the associated quiver diagram. Nodes represent vector multiplets, while the different lines encode bifundamental matter. The coloring and line style distinguish the different multiplet types, as described in Table~\ref{hypers}.

The rotation of the $(1,k)$ 5$'$ bound state relative to NS5$'$-branes modifies the original $(0,4)$ brane box model, breaking some supersymmetry and introducing mass terms for scalar fields. The extended model includes additional fundamental matter from D5$'$ and D7-branes, giving rise to new interactions in the corresponding two-dimensional field theory.

\begin{figure}[t]
	\centering
	\includegraphics[scale=0.55]{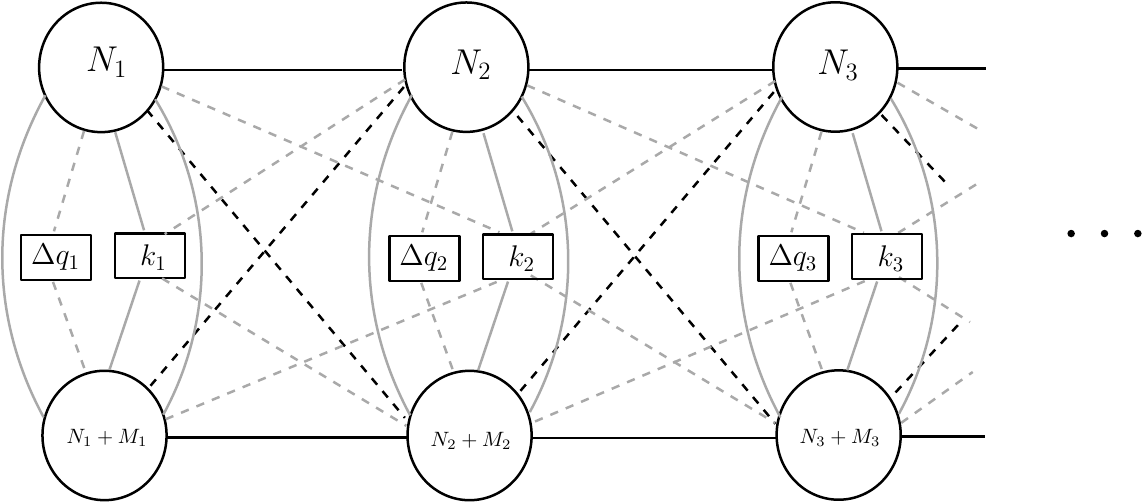}
	\caption{Quiver diagram associated with our brane setup.}
	\label{HOenhanced}
\end{figure}

Based on the field content derived from our brane box construction, we can now build the corresponding quiver gauge theory, depicted in Figure~\ref{HOenhanced}.
Due to the compactness of the $\psi$-direction, each $r$-interval contains two $(0,4)$ bifundamental hypermultiplets connecting gauge nodes of ranks $N_l$ and $N_l + M_l$, associated with D3-branes crossing NS5$'$-branes. Together, these combine into a single $(0,6)$ hypermultiplet. Additionally, two $(0,2)$ Fermi multiplets arise from open strings traversing the NS5$'$-branes at $\psi = \pi$ and $\psi = 2\pi$, which together form a $(0,4)$ Fermi multiplet connecting adjacent $r$-intervals. Each $N_l$ node also couples to $k_l$ fundamental or antifundamental $(0,4)$ hypermultiplets and to $(0,2)$ Fermi multiplets sourced by D5$'$ and D7-branes, respectively.\footnote{The D5$'$ branes contribute fundamentals to $N_l$ nodes and antifundamentals to $N_l+M_l$ nodes due to their relative positioning with respect to the NS5$'$ and $(1,k_l)$ 5-branes.}

The effect of the rotation of the $(1,k)$ 5$'$ bound state relative to the NS5$'$ branes is not yet taken into account. We begin by analyzing gauge anomaly cancellation before introducing this effect.
\begin{table}[t]
	\centering
	\begin{tabular}{|c|c|}
		\hline
		Multiplet & Contribution to gauge anomaly \\
		\hline \hline
		$(0,4)$ hyper & $+1$ \\
		$(0,4)$ vector & $-2N$ \\
		$(0,2)$ Fermi & $-1/2$ \\
		\hline
	\end{tabular}
	\caption{Gauge anomaly contributions from the multiplets in the quiver of Figure~\ref{HOenhanced}.}
	\label{Table:gaugeanom}
\end{table}
The contributions to the gauge anomaly from different multiplets are summarized in Table~\ref{Table:gaugeanom}.
Anomaly cancellation at each gauge node then requires,
\begin{equation}
	2M_l - M_{l-1} - M_{l+1} + k_l - \tfrac{1}{2}k_{l-1} - \tfrac{1}{2}k_{l+1} - \tfrac{1}{2}\Delta q_l = 0. \label{anomaly}
\end{equation}
To verify this condition, we relate the brane numbers $(N_l, M_l, k_l, q_l)$ to the quantized charges $(Q_2^l, Q_4^l, Q_6^l, Q_8^l)$ via the 'corrected dictionary',
\begin{align}
	Q_2 &= N + \tfrac{k}{12}, \nonumber\\
	Q_4 &= M - \tfrac{k}{2} + \tfrac{q}{12}, \nonumber\\
	Q_6 &= k, \nonumber\\
	Q_8 &= -q, \label{shifted}
\end{align}
as derived in \cite{Bergman:2010xd} by including Freed-Witten anomaly contributions and higher-curvature corrections.
The brane creation rules across $r$-intervals imply the recursive relations,
\begin{align}
	N_l &= N_{l-1} - M_{l-1} + k_{l-1}, \label{Seibergm2} \\
	M_l + \tfrac{q_l}{12} &= M_{l-1} + \tfrac{q_{l-1}}{12} - k_{l-1}, \label{Seibergm3} \\
	k_l &= k_{l-1} + q_{l-1}, \label{Seibergm}
\end{align}
These match the expected creation of D3, D5, and D5$'$ branes, and are consistent with 5d balancing conditions in D5$'$-D7-NS5 intersections \cite{Legramandi:2021uds}, such as,
\begin{equation}
	2k_l = k_{l-1} + k_{l+1} + \Delta q_l. \label{anomaly2}
\end{equation}
Similarly, the net change in fractional D5-brane number due to D5$'$ branes satisfies,
\begin{equation}
	2M_l = M_{l-1} + M_{l+1} - \Delta k_l, \label{anomaly3}
\end{equation}
where $\Delta k_l = k_{l-1} - k_l$ is sourced by D7-brane crossings. Substituting this into \eqref{anomaly}, we find anomaly cancellation holds only when $\Delta k_l = 0$, namely, in the massless case where $(0,4)$ supersymmetry is restored.

In the massive case, the $(1,k)$ 5$'$ bound state is rotated with respect to the NS5$'$ branes, breaking supersymmetry to $\mathcal{N}=(0,3)$ and rendering some transverse scalars massive.
 The rotation angle $\theta_l$ satisfies $\tan \theta_l = \Delta k_l$, aligning precisely with the additional term in \eqref{anomaly3}. This confirms that the $(0,3)$ quiver is anomaly-free once massive modes are properly excluded, offering a non-trivial consistency check of the supersymmetry breaking pattern.

\vspace{0.5em}

The field theory transformations induced by brane crossings,
\begin{equation}
	N \to N - M + k, \quad M \to M - k, \quad k \to k + q,
\end{equation}
were shown in \cite{Bergman:2010xd} to extend the Seiberg dualities of ABJM/ABJ \cite{Aharony:2008gk,Aharony:2009fc} to a massive, non supersymmetric setting.
 There, D8-branes act as domain walls in AdS$_4 \times \mathbb{CP}^3$, deforming ABJM to a non-supersymmetric background. The D8 backreaction preserves the AdS$_4$ geometry and interpolates to the Romans mass background \cite{Gaiotto:2009mv}. 

In our setup, the addition of NS5-branes allows a supersymmetric embedding of the D8-branes. The $r$-direction becomes holographic and drives a flow to a $(0,6)$ AdS$_3 \times \mathbb{CP}^3$ geometry, where dualities can be realized through large gauge transformations along $r$, now interpreted geometrically. This provides a new, supersymmetric extension of the Seiberg dualities proposed in \cite{Bergman:2010xd}.

\vspace{0.5em}
The holographic central charge for the $(0,6)$ SCFTs dual to the solutions with $h$-profiles given by 
\begin{equation}\label{hprofile}
	h_l(r)=Q_2^l-Q_4^l(r-l)+\frac12 Q_6^l (r-l)^2-\frac16 Q_8^l (r-l)^3,
\end{equation}
follows directly from substituting into the general formula \eqref{centralcharge}. For a suitably completed quiver, we obtain,
\begin{equation}\label{centralchargemassive}
	c_{\text{hol}} = \frac{1}{2} \sum_{l=0}^{P} \left( 2N_lk_l - M_l^2 + M_lk_l - \frac{1}{12}k_l^2 + q_l\left(N_l - \tfrac{1}{2}M_l + \tfrac{5}{12}k_l - \tfrac{13}{720}q_l\right) \right).
\end{equation}
This result provides a non-trivial prediction on the field theory side. In the massless limit ($q=0$), it reduces to the expression \eqref{centralcharge2d}, which matches the three-dimensional version \eqref{cholmassless}, previously verified through supersymmetric localization \cite{Drukker:2010nc,Herzog:2010hf,Fuji:2011km,Marino:2011eh}. Since \eqref{centralchargemassive} incorporates the shifted charges from \eqref{shifted}, it includes higher-derivative (1/N) corrections, suggesting that the corresponding field theory computation must go beyond the planar approximation.

On the field theory side, the right-moving central charge of a $(0,6)$ SCFT can be obtained from the $\mathfrak{osp}(6|2)$ superconformal algebra as \cite{Bershadsky:1986ms}
\begin{equation}\label{ccalgebra}
	c_R = \frac{k(3k+13)}{k+3},
\end{equation}
where $k$ is the level of the R-symmetry current algebra. In $(0,2)$ theories, $k$ can be computed from the 't Hooft anomaly,
\begin{equation}\label{level}
	k = \text{Tr}[\gamma_3 Q_R^2],
\end{equation}
with $Q_R$ the U$(1)_R$ charge and the trace taken over all Weyl fermions. Though our setup preserves only $(0,3)$ supersymmetry, we can still test this method in the $(0,4)$ subsector by setting $q=0$. Using the R-charges and chiralities summarized in Table~\ref{Table:R-charges}, the anomaly coefficient becomes
\begin{equation}
	k = \sum_{l=0}^{P} \left(2N_lk_l + M_lk_l - M_l^2\right),
\end{equation}
reproducing the leading terms in \eqref{centralchargemassive} up to normalization.

\begin{table}[t]
	\centering
	\begin{tabular}{|c|c|c|}
		\hline
		Multiplet & Chirality & R-charge \\
		\hline\hline
		$(0,4)$ hyper & R.H. & $-1$ \\
		$(0,4)$ twisted hyper & R.H. & $0$ \\
		$(0,4)$ vector & L.H. & $1$ \\
		$(0,2)$ Fermi & L.H. & $0$ \\
		\hline
	\end{tabular}
	\caption{R-charges and chiralities of fermions in $(0,4)$ multiplets.}
	\label{Table:R-charges}
\end{table}

The additional terms in \eqref{centralchargemassive}, proportional to $q_l$, reflect contributions from genuine $(0,3)$ multiplets. Their exact effect on the central charge remains to be understood, as no direct computation exists in the literature. This opens a promising direction for further work. One may expect mixing between the SU(2)$_R$ R-symmetry and global SU(2) symmetries, in which case the correct IR R-current --and thus the central charge-- should be obtained via $c$-extremization \cite{Benini:2012cz,Benini:2013cda}. The holographic result \eqref{centralchargemassive} thus offers a valuable benchmark for future field-theoretic studies of $(0,3)$ SCFTs.

\section{Conclusions}\label{sec:conclusions}

In this work, we have investigated a novel class of supersymmetric $\mathrm{AdS}_3$ solutions in type IIA and type IIB supergravities, preserving $\mathcal{N} = (0,6)$ and $\mathcal{N} = (0,4)$ supersymmetry, respectively. The $\mathcal{N} = (0,6)$ backgrounds are supported by an $\mathfrak{osp}(6|2)$ superconformal algebra and exhibit an enhanced $\mathrm{SO}(6)$ R-symmetry geometrically realised on a round $\mathbb{CP}^3$ parametrised as a foliation of $T^{1,1}$ over a radial direction. These solutions are completely specified by a cubic polynomial profile $h(r)$ that controls the fluxes and warping in a supersymmetry-compatible ansatz. The preserved supersymmetry is encoded in a globally defined $SU(3)$ structure on the internal manifold, derived from spinor bilinears satisfying the BPS conditions.

By performing an Abelian T-duality along the $\psi$-direction of the $T^{1,1}$, we obtained type IIB duals preserving $\mathcal{N} = (0,4)$ supersymmetry and realising an $\mathfrak{osp}(4|2)$ algebra. These backgrounds provide a more controlled setting for studying the dual two-dimensional CFTs. From them we proposed a dual two-dimensional quiver gauge theory engineered from a type IIB brane box system involving D3-, NS5-, $(1,k)$ 5$'$-, and D7-branes. The introduction of D8-branes in the type IIA frame, or, equivalently, smeared D7-branes in the dual, induces brane creation effects and modifies the ranks of the gauge groups across the quiver. Supersymmetry is reduced to $\mathcal{N} = (0,3)$ at the level of the brane intersections but is expected to enhance to $\mathcal{N} = (0,6)$ in the infrared, reflecting the enhanced supersymmetry of the dual $\mathrm{AdS}_3$ backgrounds. This structure suggests a natural interpretation of our solutions as backreacted half-BPS surface defects embedded in the ABJM/ABJ theory, realising a defect CFT via a conformal embedding of D8 and NS5 branes. Our setup thus provides a concrete and supersymmetric realisation of the proposal in~\cite{Bergman:2010gm} to use such embeddings for the construction of fractional quantum Hall edge states.

In addition, we demonstrated that large gauge transformations in the supergravity backgrounds translate into Seiberg-like dualities in the dual field theory. These transformations match precisely the rules proposed in~\cite{Bergman:2010gm}, where they were derived from ten-dimensional supergravity backgrounds dual to three-dimensional Chern-Simons matter theories. In our construction, these rules arise within a supersymmetric setting, providing a higher degree of control and enabling a holographic extension to two-dimensional $\mathcal{N} = (0,6)$ theories via dimensional reduction and defect flows. This generalises the realisation of Seiberg duality in ABJM/ABJ models to the massive case, albeit now involving a two-dimensional theory rather than a three-dimensional one.

We have also provided a concrete holographic prediction for the central charge of the dual $\mathcal{N} = (0,6)$ SCFTs, including subleading corrections arising from quantised fluxes and higher-curvature terms. Although a direct field-theoretic derivation remains elusive, particularly due to the supersymmetry reduction induced by brane rotations, our setup offers a promising framework to probe the structure of R-symmetry anomalies and central charge formulas in less explored $(0,3)$ supersymmetric theories. It would be especially interesting to relate our holographic result to exact computations of the free energy in three-dimensional $\mathcal{N} = 3$ Chern-Simons matter theories with massive deformations, as developed in~\cite{Drukker:2010nc, Herzog:2010hf, Marino:2011eh, Bergman:2009zh}. Such a comparison could provide an independent test of the proposed dimensional flow from 3d to 2d and yield novel non-perturbative insights into the strong-coupling dynamics of surface defects and their associated IR fixed points.

\appendix

\acknowledgments

We would like to thank Niall Macpherson and Nicolo Petri for collaborations in these works. YL is partially supported by the grant from the Spanish government MCIU-22PID2021-123021NB-I00. AR is supported by the Postdoctoral Fellowship Junior Marie Curie of the Research Foundation - Flanders (FWO) grant 12ACZ25N.


\begin{thebibliography}{99}
\bibitem{Strominger:1996sh}
A.~Strominger and C.~Vafa,
``Microscopic origin of the Bekenstein-Hawking entropy,''
Phys. Lett. B \textbf{379} (1996), 99-104
[arXiv:hep-th/9601029 [hep-th]].

\bibitem{Kraus:2005vz}
P.~Kraus and F.~Larsen,
``Microscopic black hole entropy in theories with higher derivatives,''
JHEP \textbf{09} (2005), 034
[arXiv:hep-th/0506176 [hep-th]].

\bibitem{Brown:1986nw}
J.~D.~Brown and M.~Henneaux,
``Central Charges in the Canonical Realization of Asymptotic Symmetries: An Example from Three-Dimensional Gravity,''
Commun. Math. Phys. \textbf{104} (1986), 207-226

\bibitem{Witten:2007kt}
E.~Witten,
``Three-Dimensional Gravity Revisited,''
[arXiv:0706.3359 [hep-th]].

\bibitem{Guica:2008mu}
M.~Guica, T.~Hartman, W.~Song and A.~Strominger,
``The Kerr/CFT Correspondence,''
Phys. Rev. D \textbf{80} (2009), 124008
[arXiv:0809.4266 [hep-th]].

\bibitem{Youm:1999zs}
D.~Youm,
``Partially localized intersecting BPS branes,''
Nucl. Phys. B \textbf{556} (1999), 222-246
[arXiv:hep-th/9902208 [hep-th]].

\bibitem{Bachas:2000fr}
C.~Bachas and M.~Petropoulos,
``Anti-de Sitter D-branes,''
JHEP \textbf{02} (2001), 025
[arXiv:hep-th/0012234 [hep-th]].

\bibitem{Gauntlett:2006af}
J.~P.~Gauntlett, O.~A.~P.~Mac Conamhna, T.~Mateos and D.~Waldram,
``Supersymmetric AdS(3) solutions of type IIB supergravity,''
Phys. Rev. Lett. \textbf{97} (2006), 171601
[arXiv:hep-th/0606221 [hep-th]].

\bibitem{Lozano:2019zvg}
Y.~Lozano, N.~T.~Macpherson, C.~Nunez and A.~Ramirez,
``Two dimensional ${\cal N}=(0,4)$ quivers dual to AdS$_3$ solutions in massive IIA,''
JHEP \textbf{01} (2020), 140
[arXiv:1909.10510 [hep-th]].

\bibitem{Couzens:2017nnr}
C.~Couzens, D.~Martelli and S.~Schafer-Nameki,
``F-theory and AdS$_{3}$/CFT$_{2}$ (2, 0),''
JHEP \textbf{06} (2018), 008
[arXiv:1712.07631 [hep-th]].

\bibitem{Faedo:2020nol}
F.~Faedo, Y.~Lozano and N.~Petri,
``Searching for surface defect CFTs within AdS$_3$,''
JHEP \textbf{11} (2020), 052
[arXiv:2007.16167 [hep-th]].

\bibitem{Lozano:2019emq}
Y.~Lozano, N.~T.~Macpherson, C.~Nunez and A.~Ramirez,
``AdS$_3$ solutions in Massive IIA with small $\mathcal{N}=(4,0)$ supersymmetry,''
JHEP \textbf{01} (2020), 129
[arXiv:1908.09851 [hep-th]].

\bibitem{Dibitetto:2018ftj}
G.~Dibitetto, G.~Lo Monaco, A.~Passias, N.~Petri and A.~Tomasiello,
``AdS$_3$ Solutions with Exceptional Supersymmetry,''
Fortsch. Phys. \textbf{66} (2018) no.10, 1800060
[arXiv:1807.06602 [hep-th]].

\bibitem{Lozano:2016wrs}
Y.~Lozano, N.~T.~Macpherson, J.~Montero and C.~Nunez,
``Three-dimensional $ \mathcal{N}=4 $ linear quivers and non-Abelian T-duals,''
JHEP \textbf{11} (2016), 133
[arXiv:1609.09061 [hep-th]].

\bibitem{Macpherson:2022sbs}
N.~T.~Macpherson and A.~Ramirez,
``AdS$_{3}$\texttimes{}S$^{2}$ in IIB with small $ \mathcal{N} $ = (4, 0) supersymmetry,''
JHEP \textbf{04} (2022), 143
[arXiv:2202.00352 [hep-th]].

\bibitem{Couzens:2021tnv}
C.~Couzens, N.~T.~Macpherson and A.~Passias,
``$ \mathcal{N} $ = (2, 2) AdS$_{3}$ from D3-branes wrapped on Riemann surfaces,''
JHEP \textbf{02} (2022), 189
[arXiv:2107.13562 [hep-th]].

\bibitem{Passias:2020ubv}
A.~Passias and D.~Prins,
``On supersymmetric AdS$_{3}$ solutions of Type II,''
JHEP \textbf{08} (2021), 168
[arXiv:2011.00008 [hep-th]].

\bibitem{Couzens:2019mkh}
C.~Couzens, H.~het Lam and K.~Mayer,
``Twisted $ \mathcal{N} $ = 1 SCFTs and their AdS$_{3}$ duals,''
JHEP \textbf{03} (2020), 032
[arXiv:1912.07605 [hep-th]].

\bibitem{Couzens:2019iog}
C.~Couzens,
``$ \mathcal{N} $ = (0, 2) AdS$_{3}$ solutions of type IIB and F-theory with generic fluxes,''
JHEP \textbf{04} (2021), 038
[arXiv:1911.04439 [hep-th]].

\bibitem{Eberhardt:2017uup}
L.~Eberhardt,
``Supersymmetric AdS$_{3}$ supergravity backgrounds and holography,''
JHEP \textbf{02} (2018), 087
[arXiv:1710.09826 [hep-th]].

\bibitem{Datta:2017ert}
S.~Datta, L.~Eberhardt and M.~R.~Gaberdiel,
``Stringy $\mathcal{N}=(2,2)$ holography for AdS${_3}$,''
JHEP \textbf{01} (2018), 146
[arXiv:1709.06393 [hep-th]].

\bibitem{Lozano:2020txg}
Y.~Lozano, C.~Nunez, A.~Ramirez and S.~Speziali,
``New AdS$_{2}$ backgrounds and $ \mathcal{N} $ = 4 conformal quantum mechanics,''
JHEP \textbf{03} (2021), 277
[arXiv:2011.00005 [hep-th]].

\bibitem{Eberhardt:2018ouy}
L.~Eberhardt, M.~R.~Gaberdiel and R.~Gopakumar,
``The Worldsheet Dual of the Symmetric Product CFT,''
JHEP \textbf{04} (2019), 103
[arXiv:1812.01007 [hep-th]].

\bibitem{Couzens:2022agr}
C.~Couzens, N.~T.~Macpherson and A.~Passias,
``On Type IIA AdS$_{3}$ solutions and massive GK geometries,''
JHEP \textbf{08} (2022), 095
[arXiv:2203.09532 [hep-th]].

\bibitem{Dibitetto:2018gtk}
G.~Dibitetto and N.~Petri,
``AdS$_{2}$ solutions and their massive IIA origin,''
JHEP \textbf{05} (2019), 107
[arXiv:1811.11572 [hep-th]].

\bibitem{Karch:2000gx}
A.~Karch and L.~Randall,
``Open and closed string interpretation of SUSY CFT's on branes with boundaries,''
JHEP \textbf{06} (2001), 063
[arXiv:hep-th/0105132 [hep-th]].

\bibitem{DeWolfe:2001pq}
O.~DeWolfe, D.~Z.~Freedman and H.~Ooguri,
``Holography and defect conformal field theories,''
Phys. Rev. D \textbf{66} (2002), 025009
[arXiv:hep-th/0111135 [hep-th]].

\bibitem{Aharony:2011yc}
O.~Aharony, L.~Berdichevsky, M.~Berkooz and I.~Shamir,
``Near-horizon solutions for D3-branes ending on 5-branes,''
Phys. Rev. D \textbf{84} (2011), 126003
[arXiv:1106.1870 [hep-th]].

\bibitem{D'Hoker:2006uv}
E.~D'Hoker, J.~Estes and M.~Gutperle,
``Interface Yang-Mills, supersymmetry, and Janus,''
Nucl. Phys. B \textbf{753} (2006), 16-41
[arXiv:hep-th/0603013 [hep-th]].

\bibitem{Chiodaroli:2011nr}
M.~Chiodaroli, E.~D'Hoker, Y.~Guo and M.~Gutperle,
``Exact half-BPS string-junction solutions in six-dimensional supergravity,''
JHEP \textbf{12} (2011), 086
[arXiv:1107.1722 [hep-th]].

\bibitem{DHoker:2007hhe}
E.~D'Hoker, J.~Estes and M.~Gutperle,
``Exact half-BPS Type IIB interface solutions. II. Flux solutions and multi-Janus,''
JHEP \textbf{06} (2007), 022
[arXiv:0705.0024 [hep-th]].

\bibitem{Billo:2016cpy}
M.~Bill\`o, V.~Gon\c{c}alves, E.~Lauria and M.~Meineri,
``Defects in conformal field theory,''
JHEP \textbf{04} (2016), 091
[arXiv:1601.02883 [hep-th]].

\bibitem{Kobayashi:2018ezm}
M.~Kobayashi, M.~Eto and M.~Nitta,
``Berezinskii-Kosterlitz-Thouless Transition of Two-Component Bose Mixtures with Intercomponent Josephson Coupling,''
Phys. Rev. Lett. \textbf{123} (2019) no.7, 075303
[arXiv:1802.08763 [cond-mat.stat-mech]].

\bibitem{DHoker:2007zhm}
E.~D'Hoker, J.~Estes and M.~Gutperle,
``Exact half-BPS Type IIB interface solutions. I. Local solution and supersymmetric Janus,''
JHEP \textbf{06} (2007), 021
[arXiv:0705.0022 [hep-th]].

\bibitem{Estes:2012nx}
J.~Estes, A.~O'Bannon, E.~Tsatis and T.~Wrase,
``Holographic Wilson Loops, Dielectric Interfaces, and Topological Insulators,''
Phys. Rev. D \textbf{87} (2013) no.10, 106005
[arXiv:1210.0534 [hep-th]].

\bibitem{Passias:2019rga}
A.~Passias and D.~Prins,
``On AdS$_3$ solutions of Type IIB,''
JHEP \textbf{05} (2020), 048
[arXiv:1910.06326 [hep-th]].

\bibitem{Macpherson:2021lbr}
N.~T.~Macpherson and A.~Tomasiello,
``$ \mathcal{N} $ = (1, 1) supersymmetric AdS$_{3}$ in 10 dimensions,''
JHEP \textbf{03} (2022), 112
[arXiv:2110.01627 [hep-th]].

\bibitem{DHoker:2008lup}
E.~D'Hoker, J.~Estes, M.~Gutperle and D.~Krym,
``Exact Half-BPS Flux Solutions in M-theory. I: Local Solutions,''
JHEP \textbf{08} (2008), 028
[arXiv:0806.0605 [hep-th]].

\bibitem{Estes:2012vm}
J.~Estes, R.~Feldman and D.~Krym,
``Exact half-BPS flux solutions in $M$ theory with D(2,1;$c^\prime$;0)$^2$ symmetry: Local solutions,''
Phys. Rev. D \textbf{87} (2013) no.4, 046008
[arXiv:1209.1845 [hep-th]].

\bibitem{Bachas:2013vza}
C.~Bachas, E.~D'Hoker, J.~Estes and D.~Krym,
``M-theory Solutions Invariant under $D(2,1;\gamma) \oplus D(2,1;\gamma)$,''
Fortsch. Phys. \textbf{62} (2014), 207-254
[arXiv:1312.5477 [hep-th]].

\bibitem{Macpherson:2018mif}
N.~T.~Macpherson,
``Type II solutions on AdS$_{3} \times$ S$^{3} \times$ S$^{3}$ with large superconformal symmetry,''
JHEP \textbf{05} (2019), 089
[arXiv:1812.10172 [hep-th]].

\bibitem{Legramandi:2020txf}
A.~Legramandi, G.~Lo Monaco and N.~T.~Macpherson,
``All $\mathcal{N}=(8,0)$ AdS$_3$ solutions in 10 and 11 dimensions,''
JHEP \textbf{05} (2021), 263
[arXiv:2012.10507 [hep-th]].


\bibitem{Dibitetto:2020bsh}
G.~Dibitetto and N.~Petri,
``AdS$_{3}$ from M-branes at conical singularities,''
JHEP \textbf{01} (2021), 129
[arXiv:2010.12323 [hep-th]].

\bibitem{Gauntlett:2007ph}
J.~P.~Gauntlett and O.~A.~P.~Mac Conamhna,
``AdS spacetimes from wrapped D3-branes,''
Class. Quant. Grav. \textbf{24} (2007), 6267-6286
[arXiv:0707.3105 [hep-th]].

\bibitem{Conti:2024rwd}
A.~Conti, G.~Dibitetto, Y.~Lozano, N.~Petri and A.~Ram\'\i{}rez,
``Half-BPS Janus solutions in AdS$_{7}$,''
JHEP \textbf{12} (2024), 198
[arXiv:2407.21619 [hep-th]].

\bibitem{Lozano:2022ouq}
Y.~Lozano, N.~T.~Macpherson, N.~Petri and C.~Risco,
``New AdS$_{3}$/CFT$_{2}$ pairs in massive IIA with (0, 4) and (4, 4) supersymmetries,''
JHEP \textbf{09} (2022), 130
[arXiv:2206.13541 [hep-th]].

\bibitem{Capuozzo:2024onf}
P.~Capuozzo, J.~Estes, B.~Robinson and B.~Suzzoni,
``From large to small $ \mathcal{N} $ = (4, 4) superconformal surface defects in holographic 6d SCFTs,''
JHEP \textbf{08} (2024), 094
[arXiv:2402.11745 [hep-th]].

\bibitem{Conti:2024rqy}
A.~Conti and N.~T.~Macpherson,
``${\cal N}=(4,4)$ supersymmetric AdS$_3$ solutions in $d=11$,''
[arXiv:2408.17303 [hep-th]].

\bibitem{Kim:2005ez}
N.~Kim,
``AdS(3) solutions of IIB supergravity from D3-branes,''
JHEP \textbf{01} (2006), 094
[arXiv:hep-th/0511029 [hep-th]].

\bibitem{Ramirez:2025jut}
A.~Ramirez and S.~Zacarias,
``On deformations of AdS$_3$ solutions, supersymmetry and $G$-structures,''
[arXiv:2504.11207 [hep-th]].

\bibitem{Lozano:2015bra}
Y.~Lozano, N.~T.~Macpherson, J.~Montero and E.~\'O.~Colg\'ain,
``New $AdS_3 \times S^2$ T-duals with $ \mathcal{N}=\left(0,4\right) $ supersymmetry,''
JHEP \textbf{08} (2015), 121
[arXiv:1507.02659 [hep-th]].

\bibitem{Couzens:2017way}
C.~Couzens, C.~Lawrie, D.~Martelli, S.~Schafer-Nameki and J.~M.~Wong,
``F-theory and AdS$_{3}$/CFT$_{2}$,''
JHEP \textbf{08} (2017), 043
[arXiv:1705.04679 [hep-th]].

\bibitem{Lozano:2020bxo}
Y.~Lozano, C.~Nunez, A.~Ramirez and S.~Speziali,
``$M$-strings and AdS$_3$ solutions to M-theory with small $\mathcal{N}=(0,4)$ supersymmetry,''
JHEP \textbf{08} (2020), 118
[arXiv:2005.06561 [hep-th]].

\bibitem{Faedo:2020lyw}
F.~Faedo, Y.~Lozano and N.~Petri,
``New $\mathcal{N}=(0,4)$ AdS$_3$ near-horizons in Type IIB,''
JHEP \textbf{04} (2021), 028
[arXiv:2012.07148 [hep-th]].

\bibitem{Tong:2014yna}
D.~Tong,
``The holographic dual of $AdS_{3} \times  S^{3} \times S^{3} \times S^{1}$,''
JHEP \textbf{04} (2014), 193
[arXiv:1402.5135 [hep-th]].

\bibitem{Lozano:2015cra}
Y.~Lozano, N.~T.~Macpherson and J.~Montero,
``A $ \mathcal{N}=2 $ supersymmetric AdS$_{4}$ solution in M-theory with purely magnetic flux,''
JHEP \textbf{10} (2015), 004
[arXiv:1507.02660 [hep-th]].



\bibitem{Kelekci:2016uqv}
\"O.~Kelekci, Y.~Lozano, J.~Montero, E.~\'O.~Colg\'ain and M.~Park,
``Large superconformal near-horizons from M-theory,''
Phys. Rev. D \textbf{93} (2016) no.8, 086010
[arXiv:1602.02802 [hep-th]].

\bibitem{Eberhardt:2017pty}
L.~Eberhardt, M.~R.~Gaberdiel and W.~Li,
``A holographic dual for string theory on AdS$_{3}$\texttimes{}S$^{3}$\texttimes{}S$^{3}$\texttimes{}S$^{1}$,''
JHEP \textbf{08} (2017), 111
[arXiv:1707.02705 [hep-th]].

\bibitem{Dibitetto:2017tve}
G.~Dibitetto and N.~Petri,
``BPS objects in D = 7 supergravity and their M-theory origin,''
JHEP \textbf{12} (2017), 041
[arXiv:1707.06152 [hep-th]].

\bibitem{Dibitetto:2017klx}
G.~Dibitetto and N.~Petri,
``6d surface defects from massive type IIA,''
JHEP \textbf{01} (2018), 039
[arXiv:1707.06154 [hep-th]].

\bibitem{Gaberdiel:2018rqv}
M.~R.~Gaberdiel and R.~Gopakumar,
``Tensionless string spectra on AdS$_{3}$,''
JHEP \textbf{05} (2018), 085
[arXiv:1803.04423 [hep-th]].

\bibitem{Eberhardt:2018sce}
L.~Eberhardt and I.~G.~Zadeh,
``$\mathcal{N}=(3,3)$ holography on ${\rm AdS}_3 \times ({\rm S}^3 \times {\rm S}^3 \times {\rm S}^1)/\mathbb Z_2$,''
JHEP \textbf{07} (2018), 143
[arXiv:1805.09832 [hep-th]].

\bibitem{Dibitetto:2018iar}
G.~Dibitetto and N.~Petri,
``Surface defects in the D4 $-$ D8 brane system,''
JHEP \textbf{01} (2019), 193
[arXiv:1807.07768 [hep-th]].

\bibitem{Lozano:2019jza}
Y.~Lozano, N.~T.~Macpherson, C.~Nunez and A.~Ramirez,
``1/4 BPS solutions and the AdS$_3$/CFT$_2$ correspondence,''
Phys. Rev. D \textbf{101} (2020) no.2, 026014
[arXiv:1909.09636 [hep-th]].

\bibitem{Lozano:2019ywa}
Y.~Lozano, N.~T.~Macpherson, C.~Nunez and A.~Ramirez,
``AdS$_3$ solutions in massive IIA, defect CFTs and T-duality,''
JHEP \textbf{12} (2019), 013
[arXiv:1909.11669 [hep-th]].

\bibitem{Eberhardt:2019ywk}
L.~Eberhardt, M.~R.~Gaberdiel and R.~Gopakumar,
``Deriving the AdS$_{3}$/CFT$_{2}$ correspondence,''
JHEP \textbf{02} (2020), 136
[arXiv:1911.00378 [hep-th]].

\bibitem{Legramandi:2019xqd}
A.~Legramandi and N.~T.~Macpherson,
``AdS$_3$ solutions with from $\mathcal{N}=(3,0)$ from S$^3\times$S$^3$  fibrations,''
Fortsch. Phys. \textbf{68} (2020) no.3-4, 2000014
[arXiv:1912.10509 [hep-th]].


\bibitem{Couzens:2021veb}
C.~Couzens, Y.~Lozano, N.~Petri and S.~Vandoren,
``N=(0,4) black string chains,''
Phys. Rev. D \textbf{105} (2022) no.8, 086015
[arXiv:2109.10413 [hep-th]].

\bibitem{Macpherson:2023cbl}
N.~T.~Macpherson and A.~Ramirez,
``AdS$_{3}$ vacua realising $ \mathfrak{osp} $(n|2) superconformal symmetry,''
JHEP \textbf{08} (2023), 024
[arXiv:2304.12207 [hep-th]].

\bibitem{Lozano:2024idt}
Y.~Lozano, N.~T.~Macpherson, N.~Petri and A.~Ram\'\i{}rez,
``Holographic $ \frac{1}{2} $-BPS surface defects in ABJM,''
JHEP \textbf{08} (2024), 044
[arXiv:2404.17469 [hep-th]].

\bibitem{Conti:2025djz}
A.~Conti, Y.~Lozano and N.~T.~Macpherson,
``$\mathcal{N}=6$ supersymmetric AdS$_2 \times \mathbb{CP}^3\times \Sigma_2 $,''
[arXiv:2503.23585 [hep-th]].



\bibitem{Gauntlett:1997pk}
J.~P.~Gauntlett, G.~W.~Gibbons, G.~Papadopoulos and P.~K.~Townsend,
``Hyper-Kahler manifolds and multiply intersecting branes,''
Nucl. Phys. B \textbf{500} (1997), 133-162
[arXiv:hep-th/9702202 [hep-th]].

\bibitem{Klebanov:2007ws}
I.~R.~Klebanov, D.~Kutasov and A.~Murugan,
``Entanglement as a probe of confinement,''
Nucl. Phys. B \textbf{796} (2008), 274-293
[arXiv:0709.2140 [hep-th]].

\bibitem{Macpherson:2014eza}
N.~T.~Macpherson, C.~N\'u\~nez, L.~A.~Pando Zayas, V.~G.~J.~Rodgers and C.~A.~Whiting,
``Type IIB supergravity solutions with AdS$_{5}$ from Abelian and non-Abelian T dualities,''
JHEP \textbf{02} (2015), 040
[arXiv:1410.2650 [hep-th]].

\bibitem{Bea:2015fja}
Y.~Bea, J.~D.~Edelstein, G.~Itsios, K.~S.~Kooner, C.~Nunez, D.~Schofield and J.~A.~Sierra-Garcia,
``Compactifications of the Klebanov-Witten CFT and new AdS$_{3}$ backgrounds,''
JHEP \textbf{05} (2015), 062
[arXiv:1503.07527 [hep-th]].

\bibitem{Aharony:2008gk}
O.~Aharony, O.~Bergman and D.~L.~Jafferis,
``Fractional M2-branes,''
JHEP \textbf{11} (2008), 043
[arXiv:0807.4924 [hep-th]].

\bibitem{Aharony:2008ug}
O.~Aharony, O.~Bergman, D.~L.~Jafferis and J.~Maldacena,
``N=6 superconformal Chern-Simons-matter theories, M2-branes and their gravity duals,''
JHEP \textbf{10} (2008), 091
[arXiv:0806.1218 [hep-th]].

\bibitem{Aharony:2009fc}
O.~Aharony, A.~Hashimoto, S.~Hirano and P.~Ouyang,
``D-brane Charges in Gravitational Duals of 2+1 Dimensional Gauge Theories and Duality Cascades,''
JHEP \textbf{01} (2010), 072
[arXiv:0906.2390 [hep-th]].

\bibitem{Freed:1999vc}
D.~S.~Freed and E.~Witten,
``Anomalies in string theory with D-branes,''
Asian J. Math. \textbf{3} (1999), 819
[arXiv:hep-th/9907189 [hep-th]].

\bibitem{Bergman:2009zh}
O.~Bergman and S.~Hirano,
``Anomalous radius shift in AdS(4)/CFT(3),''
JHEP \textbf{07} (2009), 016
[arXiv:0902.1743 [hep-th]].

\bibitem{Bergman:2013qoa}
O.~Bergman, S.~Hirano and G.~Lifschytz,
``Some New Results in $AdS_{4}/CFT_{3}$ Duality,''
J. Phys. Conf. Ser. \textbf{462} (2013) no.1, 012003

\bibitem{Drukker:2010nc}
N.~Drukker, M.~Marino and P.~Putrov,
``From weak to strong coupling in ABJM theory,''
Commun. Math. Phys. \textbf{306} (2011), 511-563
[arXiv:1007.3837 [hep-th]].

\bibitem{Herzog:2010hf}
C.~P.~Herzog, I.~R.~Klebanov, S.~S.~Pufu and T.~Tesileanu,
``Multi-Matrix Models and Tri-Sasaki Einstein Spaces,''
Phys. Rev. D \textbf{83} (2011), 046001
[arXiv:1011.5487 [hep-th]].

\bibitem{Fuji:2011km}
H.~Fuji, S.~Hirano and S.~Moriyama,
``Summing Up All Genus Free Energy of ABJM Matrix Model,''
JHEP \textbf{08} (2011), 001
[arXiv:1106.4631 [hep-th]].

\bibitem{Marino:2011eh}
M.~Marino and P.~Putrov,
``ABJM theory as a Fermi gas,''
J. Stat. Mech. \textbf{1203} (2012), P03001
[arXiv:1110.4066 [hep-th]].

\bibitem{Hanany:2018hlz}
A.~Hanany and T.~Okazaki,
``(0,4) brane box models,''
JHEP \textbf{03} (2019), 027
[arXiv:1811.09117 [hep-th]].

\bibitem{Kitao:1998mf}
T.~Kitao, K.~Ohta and N.~Ohta,
``Three-dimensional gauge dynamics from brane configurations with (p,q)-fivebrane,''
Nucl. Phys. B \textbf{539} (1999), 79-106
[arXiv:hep-th/9808111 [hep-th]].

\bibitem{Bergman:1999na}
O.~Bergman, A.~Hanany, A.~Karch and B.~Kol,
``Branes and supersymmetry breaking in three-dimensional gauge theories,''
JHEP \textbf{10} (1999), 036
[arXiv:hep-th/9908075 [hep-th]].

\bibitem{Bergman:2010xd}
O.~Bergman and G.~Lifschytz,
``Branes and massive IIA duals of 3d CFT's,''
JHEP \textbf{04} (2010), 114
[arXiv:1001.0394 [hep-th]].

\bibitem{Legramandi:2021uds}
A.~Legramandi and C.~Nunez,
``Electrostatic description of five-dimensional SCFTs,''
Nucl. Phys. B \textbf{974} (2022), 115630
[arXiv:2104.11240 [hep-th]].

\bibitem{Gaiotto:2009mv}
D.~Gaiotto and A.~Tomasiello,
``The gauge dual of Romans mass,''
JHEP \textbf{01} (2010), 015
[arXiv:0901.0969 [hep-th]].

\bibitem{Bershadsky:1986ms}
M.~A.~Bershadsky,
``Superconformal Algebras in Two-dimensions With Arbitrary $N$,''
Phys. Lett. B \textbf{174} (1986), 285-288

\bibitem{Benini:2012cz}
F.~Benini and N.~Bobev,
``Exact two-dimensional superconformal R-symmetry and c-extremization,''
Phys. Rev. Lett. \textbf{110} (2013) no.6, 061601
[arXiv:1211.4030 [hep-th]].

\bibitem{Benini:2013cda}
F.~Benini and N.~Bobev,
``Two-dimensional SCFTs from wrapped branes and c-extremization,''
JHEP \textbf{06} (2013), 005
[arXiv:1302.4451 [hep-th]].

\bibitem{Bergman:2010gm}
O.~Bergman, N.~Jokela, G.~Lifschytz and M.~Lippert,
``Quantum Hall Effect in a Holographic Model,''
JHEP \textbf{10} (2010), 063
[arXiv:1003.4965 [hep-th]].

\end{thebibliography}
\end{document}